\documentclass[preprintnumbers,amsmath,amssymb]{revtex4}

\usepackage{epsfig}
\usepackage{graphicx}
\usepackage{color} 

\usepackage{fancyheadings} 
\fancypagestyle{first}{%
\fancyhf{} 
\cfoot{
Approved for public release; unlimited distribution. \copyright 2017 BAE Systems. All rights reserved.
} 
 
} 

\newcommand{\tstamp}{\today}   
\makeatletter
\newcommand{\xMapsto}[2][]{\ext@arrow 0599{\Mapstofill@}{#1}{#2}}
\def\Mapstofill@{\arrowfill@{\Mapstochar\Relbar}\Relbar\Rightarrow}
\makeatother

\begin{document}

\title{Topological density estimation}

\author{Steve Huntsman}%
\email{steve.huntsman@baesystems.com}
\affiliation{BAE Systems, 4301 North Fairfax Drive, Arlington, Virginia 22203
}%

\date{\today}

\begin{abstract}
We introduce \emph{topological density estimation} (TDE), in which the multimodal structure of a probability density function is topologically inferred and subsequently used to perform bandwidth selection for kernel density estimation. We show that TDE has performance and runtime advantages over competing methods of kernel density estimation for highly multimodal probability density functions. We also show that TDE yields useful auxiliary information, that it can determine its own suitability for use, and we explain its performance. 
\end{abstract}


\maketitle

\thispagestyle{first}


\section{\label{sec:Introduction}Introduction}

Kernel density estimation (KDE) is such a ubiquitous and fundamental technique in statistics that our claim in this paper of an interesting and useful new contribution to the enormous body of literature (see, e.g., \cite{DevroyeLugosi,Tsybakov,Wasserman,ZambomDias}) almost inevitably entails some degree of hubris. Even the idea of using KDE to determine the multimodal structure of a probability density function (henceforth simply called ``density'') has by now a long history \cite{Silverman,Minnotte,HallMinnotteZhang} that has very recently been explicitly coupled with ideas of topological persistence \cite{BauerEtAl}. 

In this paper, we take precisely the opposite course and use the multimodal structure of a density to perform bandwidth selection for KDE, an approach we call \emph{topological density estimation} (TDE). The paper is organized as follows: in \S \ref{sec:TopologicalDensityEstimation}, we outline TDE via the enabling construction of unimodal category and the corresponding decomposition detailed in \cite{BaryshnikovGhrist}. In \S \ref{sec:Evaluation}, we evaluate TDE along the same lines as \cite{HeidenreichSchindlerSperlich} and show that it offers advantages over its competitors for highly multimodal densities, despite requiring no parameters or nontrivial configuration choices. Finally, in \S \ref{sec:Remarks} we make some remarks on TDE. Scripts and code used to produce our results are included in appendices, as are additional figures.

Surprisingly, our simple idea of combining the (already topological) notion of unimodal category with ideas of topological persistence \cite{Ghrist,Oudot} has not hitherto been considered in the literature, though the related idea of combining multiresolution analysis with KDE is well-established \cite{ChaudhuriMarron}. The work closest in spirit to ours appears to be \cite{PokornyEtAl1} (see also \cite{PokornyEtAl2}), in which the idea of using persistent homology to simultaneously estimate the support of a compactly supported (typically multivariate) density and a bandwidth for a compact kernel was explored. In particular, our work also takes the approach of simultaneously estimating a topological datum and selecting a kernel bandwidth in a mutually reinforcing way. Furthermore, in the particular case where a density is a convex combination of widely separated unimodal densities, their constructions and ours will manifest some similarity, and in general using a kernel with compact support would allow the techniques of \cite{PokornyEtAl1} and the present paper to be used in concert. However, our technique is in most ways much simpler and KDE in one dimension typically features situations in which the support of the underlying density is or can be assumed to be topologically trivial, so we do not explore this integration here.

\section{\label{sec:TopologicalDensityEstimation}Topological density estimation}

\begin{figure}[htbp]
\includegraphics[trim = 10mm 80mm 10mm 5mm, clip, width=120mm, keepaspectratio]{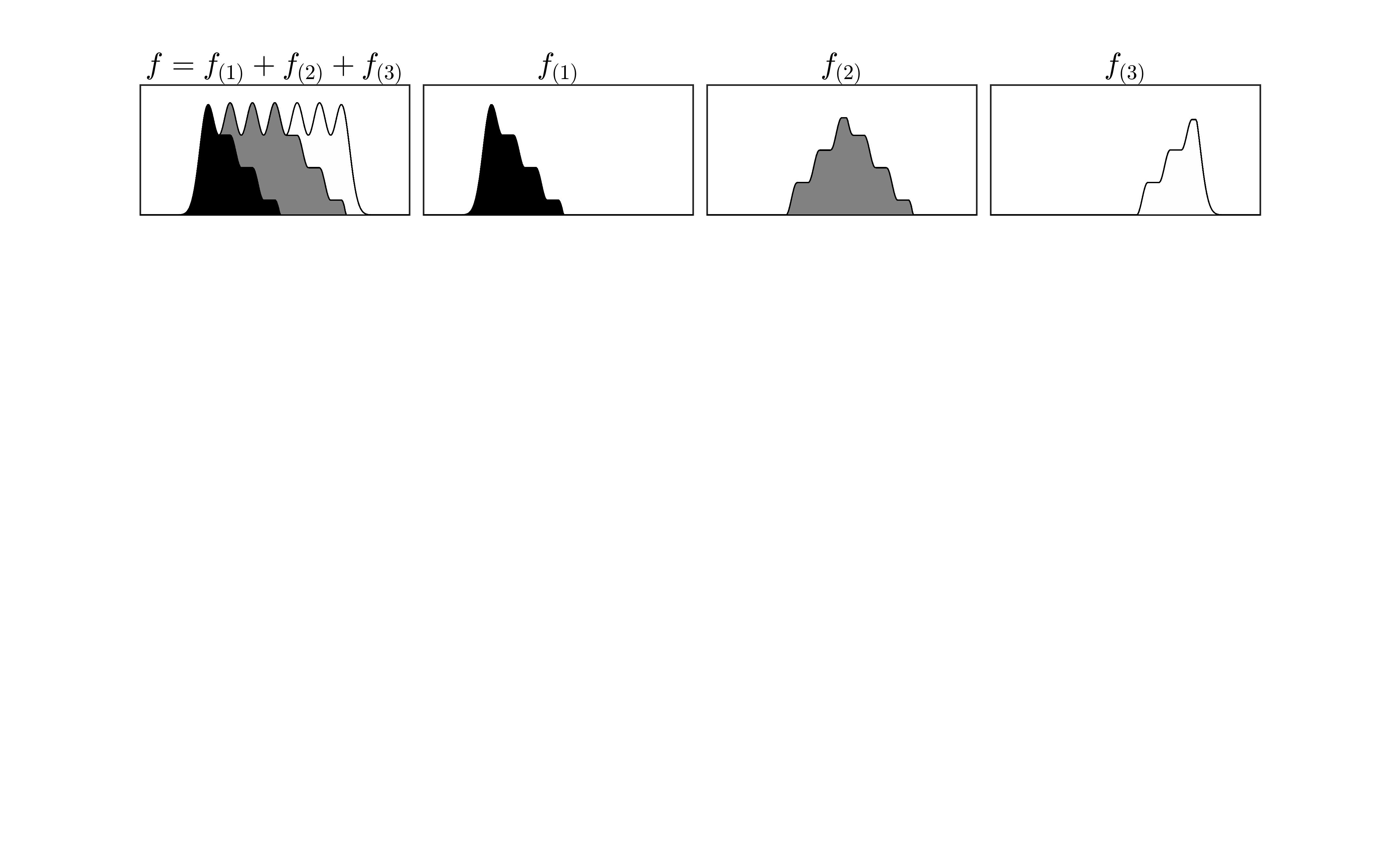}
\caption{ \label{fig:unidec} Example of an explicit unimodal decomposition obtained using the ``sweep'' algorithm implemented in \S \ref{sec:unidec}.} 
\end{figure}

Let $\mathcal{D}(\mathbb{R}^d)$ denote the space of continuous densities on $\mathbb{R}^d$. $f \in \mathcal{D}(\mathbb{R}^d)$ is \emph{unimodal} if $f^{-1}([y,\infty))$ is contractible (within itself) for all $y$. The \emph{unimodal category} $\text{ucat}(f)$ is the least integer such that there exist unimodal densities $f_j \in \mathcal{D}(\mathbb{R}^d)$ for $1 \le j \le \text{ucat}(f)$ and $0 < c_j$ with $\sum_j c_j = 1$ and $f = \sum_j c_j f_j$: we call the RHS a \emph{unimodal decomposition} of $f$ (see figure \ref{fig:unidec}). That is, $\text{ucat}(f)$ is the minimal number of unimodal components whose convex combination yields $f$. For example, in practice the unimodal category of a Gaussian mixture model is usually (but not necessarily) the number of components.

	%
	%
	%

Note that while a unimodal decomposition is very far from unique, the unimodal category is a topological invariant. For, let $g : \mathbb{R}^d \rightarrow \mathbb{R}^d$ be a homeomorphism: then since $g((f \circ g)^{-1}([y,\infty))) = f^{-1}([y,\infty))$, it follows that $f$ is unimodal iff $f \circ g$ is. In situations where there is no preferred coordinate system (such as in, e.g., certain problems of distributed sensing), the analytic details of a density are irrelevant, whereas the topologically invariant features such as the unimodal category are essential. 

The essential idea of TDE is this: given a kernel $K(x)$ and sample data $X_j$ for $1 \le j \le n$, for each proposed bandwidth $h$, compute the density estimate
\begin{equation}
\label{eq:KDE}
\hat f_{h;X}(x) := \frac{1}{nh} \sum_{j=1}^n K \left ( \frac{x-X_j}{h} \right )
\end{equation}
and subsequently 
\begin{equation}
\label{eq:ucatKDE}
u_X(h) := \text{ucat}(\hat f_{h;X}).
\end{equation}
The concomitant estimate of the unimodal category is
\begin{equation}
\label{eq:ucatEstimate}
\hat m := \arg \max_m \mu(u_X^{-1}(m))
\end{equation}
where $\mu$ denotes an appropriate measure (nominally Lebesgue measure). Now $u_X^{-1}(\hat m)$ is the set of bandwidths consistent with the estimated unimodal category. TDE amounts to choosing the bandwidth
\begin{equation}
\label{eq:TDE}
\hat h := \text{median}_\mu ( u_X^{-1}(\hat m) ).
\end{equation}

That is, we look for the largest set $u_X^{-1}(\hat m)$ where $u_X$ is constant--i.e., where the value of the unimodal category is the most prevalent (and usually in practice, also \emph{persistent})--and pick as bandwidth the central element of $u_X^{-1}(\hat m)$.

\section{\label{sec:Evaluation}Evaluation}

In this section, we evaluate the performance of TDE and compare it to other methods following \cite{HeidenreichSchindlerSperlich}.

\subsection{\label{sec:Protocol}Protocol}

Write
\begin{equation}
\label{eq:Gamma}
Gamma(x;k,\theta) := \frac{1}{\Gamma(k) \theta} (x/\theta)^{k-1} e^{-x/\theta} \cdot 1_{\mathbb{R}_{\ge 0}}(x)
\end{equation}
and
\begin{equation}
\label{eq:Normal}
Normal(x;\mu,\sigma) := \frac{1}{\sqrt{2 \pi \sigma^2}} e^{-(x-\mu)^2/2\sigma^2}.
\end{equation}

\begin{figure}[htbp]
\includegraphics[trim = 30mm 170mm 30mm 20mm, clip, width=180mm,keepaspectratio]{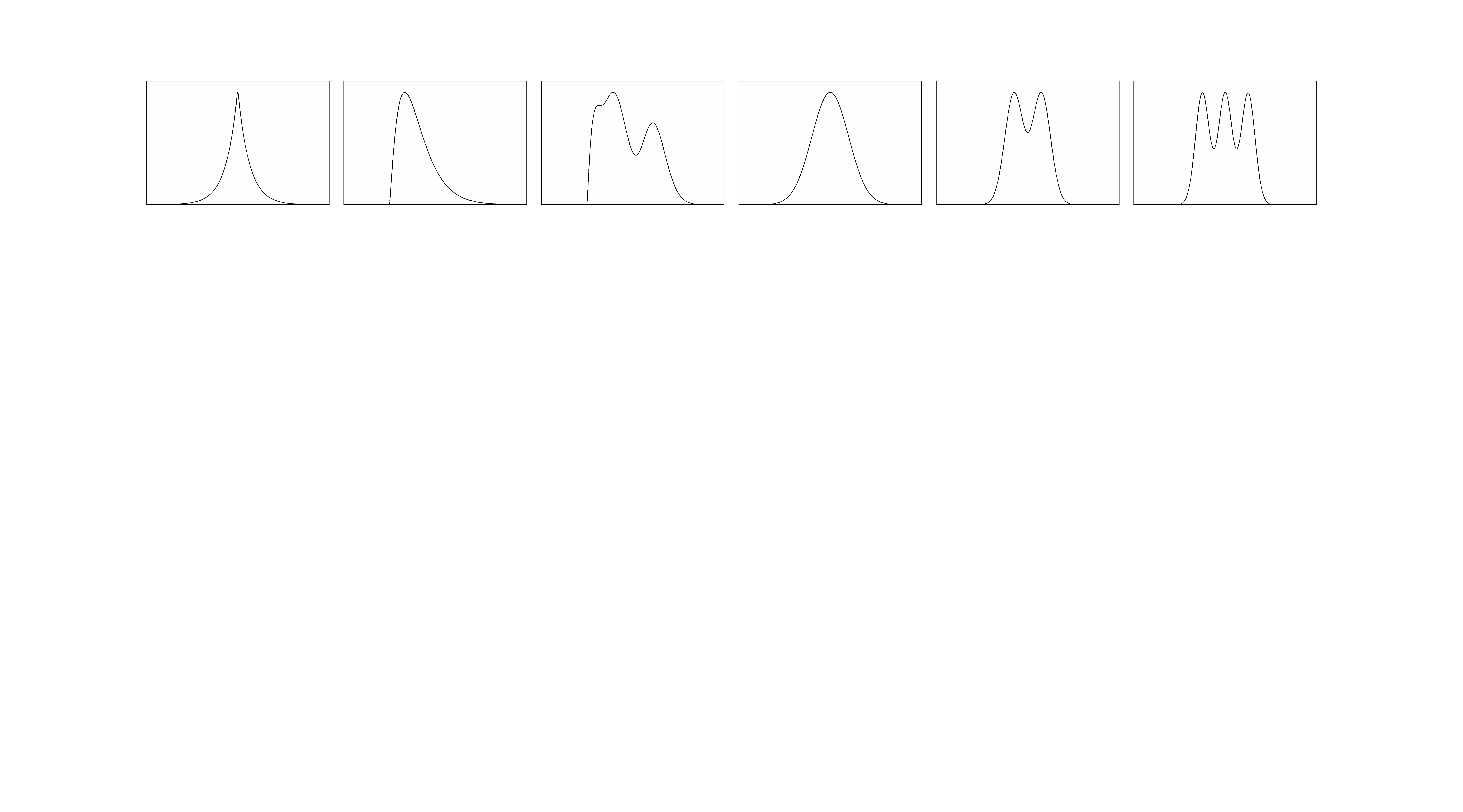}
\caption{ \label{fig:f1-6} (L-R) The six densities $f_1, \dots, f_6$ in \eqref{eq:f_j} and from \cite{HeidenreichSchindlerSperlich} over the interval $[-0.5,1.5]$.} 
\end{figure}

We estimate the same six densities as in \cite{HeidenreichSchindlerSperlich} (see figure \ref{fig:f1-6}), viz.
\begin{subequations}
\label{eq:f_j}
\begin{align}
f_1(x) \quad := 
	& \quad \frac{1}{2b_1} e^{-|x-\mu_1|/b_1}, \quad \mu_1 = 0.5, \quad b_1 = 0.125; \label{eq:f1} \\
f_2(x) \quad := 
	& \quad Gamma(x;b_2^2,1/\lambda_2 b_2), \quad b_2 = 1.5, \quad \lambda_2 = 5; \label{eq:f2} \\
f_3(x) \quad := 
	& \quad \frac{1}{3} \sum_{j = 1}^3 Gamma(x;b_{3j}^2,1/\lambda_3 b_{3j}), \quad b_{31} = 1.5, \quad b_{32} = 3, \quad b_{33} = 6, \quad \lambda_3 = 8; \label{eq:f3} \\
f_4(x) \quad := 
	& \quad Normal(x;\mu_4,\sigma_4), \quad \mu_4 = 0.5, \quad \sigma_4 = 0.2; \label{eq:f4} \\
f_5(x) \quad := 
	& \quad \frac{1}{2} \sum_{j = 1}^2 Normal(x;\mu_{5j},\sigma_5), \quad \mu_{51} = 0.35, \quad \mu_{52} = 0.65, \quad \sigma_{5} = 0.1; \label{eq:f5} \\
f_6(x) \quad := 
	& \quad \frac{1}{3} \sum_{j = 1}^3 Normal(x;\mu_{6j},\sigma_6), \quad \mu_{61} = 0.25, \quad \mu_{62} = 0.5, \quad \mu_{63} = 0.75, \quad \sigma_6 = 0.075. \label{eq:f6}
\end{align}
\end{subequations}

\begin{figure}[htbp]
\includegraphics[trim = 30mm 95mm 30mm 20mm, clip, width=180mm,keepaspectratio]{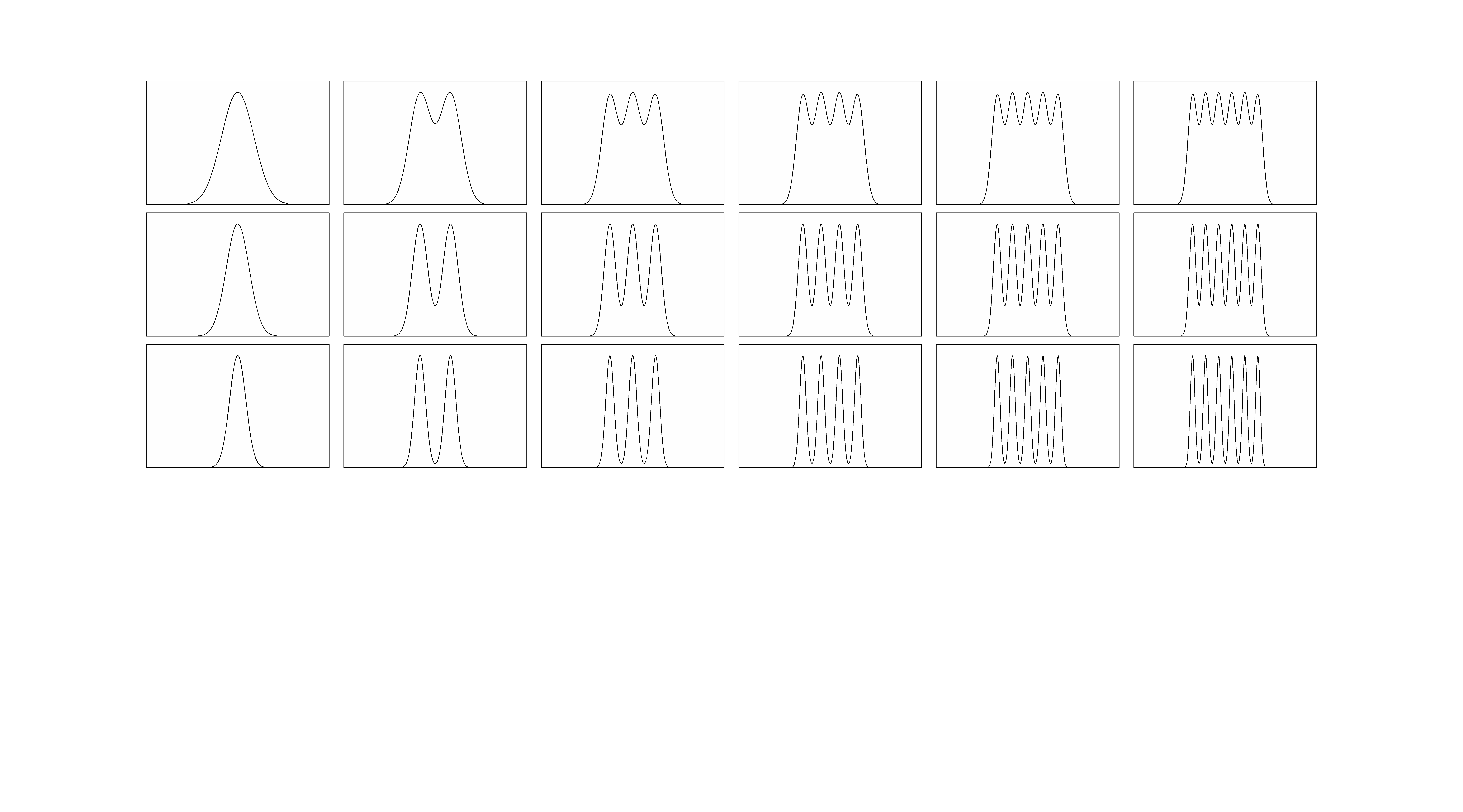}
\caption{ \label{fig:fkm} The densities $f_{km}$ for $1 \le k \le 3$ and $1 \le m \le 6$ over the interval $[-0.5,1.5]$. Rows are indexed by $k$ and columns are indexed by $m$; the upper left panel shows $f_{11}$ and the lower right panel shows $f_{36}$. 
} 
\end{figure}

However, in the present context it is also particularly relevant to estimate highly multimodal densities. 
\footnote{
The analysis of \cite{HeidenreichSchindlerSperlich} ``excludes [functions with] sharp peaks and highly oscillating functions [because they] should not be tackled with kernels anyway.'' We feel that this reasoning is debatable in light of the qualitative performance and runtime advantages of TDE on highly multimodal densities with a number of samples sufficient to plausibly permit good estimates.
}
Towards that end we also consider the following family:
\begin{equation}
\label{eq:fkm}
f_{km}(x) := \frac{1}{m} \sum_{j = 1}^m Normal \left ( x;\frac{j}{m+1},\frac{1}{2^{k+2}(m+1)^2} \right )
\end{equation}
for $1 \le k \le 3$ and $1 \le m \le 10$ (see figure \ref{fig:fkm}). 

We also consider performance measures strictly generalizing the five below used by \cite{HeidenreichSchindlerSperlich}, viz.
\begin{itemize}
\item $c_1 := \text{mean}(\hat h - h_{opt})$;
\item $c_2 := \text{mean}(\text{ISE}(\hat h))$;
\item $c_3 := \text{std}(\text{ISE}(\hat h))$;
\item $c_4 := \text{mean}([\text{ISE}(\hat h)-\text{ISE}(h_{opt})]^2)$;
\item $c_5 := \text{mean}(|\text{ISE}(\hat h)-\text{ISE}(h_{opt})|)$.
\end{itemize}
Here the \emph{integrated squared error} is $\text{ISE}(h) := \| \hat f_{h;X} - f \|_2^2$ and $h_{opt} := \arg \min_h \text{ISE}(h)$ (note that $\text{ISE}(h)$ is sample-dependent, whereby $h_{opt}$ is also). Specifically, we consider the \emph{distributions} of
\begin{itemize}
\item $\hat h - h_{opt}$ rather than its mean $c_1$;
\item $\text{ISE}(\hat h)$ rather than its mean $c_2$ and standard deviation $c_3$;
\item $c_{45} := \log_{10} | \text{ISE}(\hat h)-\text{ISE}(h_{opt}) |$ rather than $c_4$ and $c_5$ (note that $c_4 = \text{mean}(10^{2c_{45}})$ and $c_5 = \text{mean}(10^{c_{45}})$).
\end{itemize}
We will illustrate the distributions \emph{in toto} and thereby obtain more information than from the summary statistics $c_j$ by careful use of pseudotransparency plots.

As in \cite{HeidenreichSchindlerSperlich}, all results below are based on $N = 250$ simulation runs and sample sizes $n \in \{25,50,100,200,500\}$, noting that values $n \in \{500,1000\}$ were evaluated but not explicitly shown in the results of \cite{HeidenreichSchindlerSperlich}. We consider both Gaussian and Epanechnikov kernels, and it turns out to be broadly sufficient to consider just TDE and ordinary least-squares cross-validation KDE (CV). 

Before discussing the results, we finally note that it is necessary to deviate slightly from the evaluation protocol of \cite{HeidenreichSchindlerSperlich} in one respect that is operationally insignificant but conceptually essential. Because TDE hinges on identifying a persistent unimodal category, selecting bandwidths from the sparse set of 25 logarithmically spaced points from $n^{-1}$ to $1$ used across methods in \cite{HeidenreichSchindlerSperlich} is fundamentally inappropriate for evaluating the potential of TDE. Instead, we use (for both CV and TDE) the general data-adaptive bandwidth set $\{\Delta X / j : 1 \le j \le n\}$, where $\Delta X := \max(X) - \min(X)$ and $X$ denotes the sampled data.
\footnote{
In applications, a sensible choice of bandwidth set would be something more like $\{\Delta X / j : 1 \le j \le \min(n,100)\}$, with details determined (as usual) by the problem at hand. While it may be reasonable to dispense with constant spacing in a bandwidth set, it is absolutely essential to have enough members of the set to give persistent results.
}

\subsection{\label{sec:Results}Results}

\begin{figure}[htbp]
\includegraphics[trim = 35mm 15mm 25mm 10mm, clip, width=180mm,keepaspectratio]{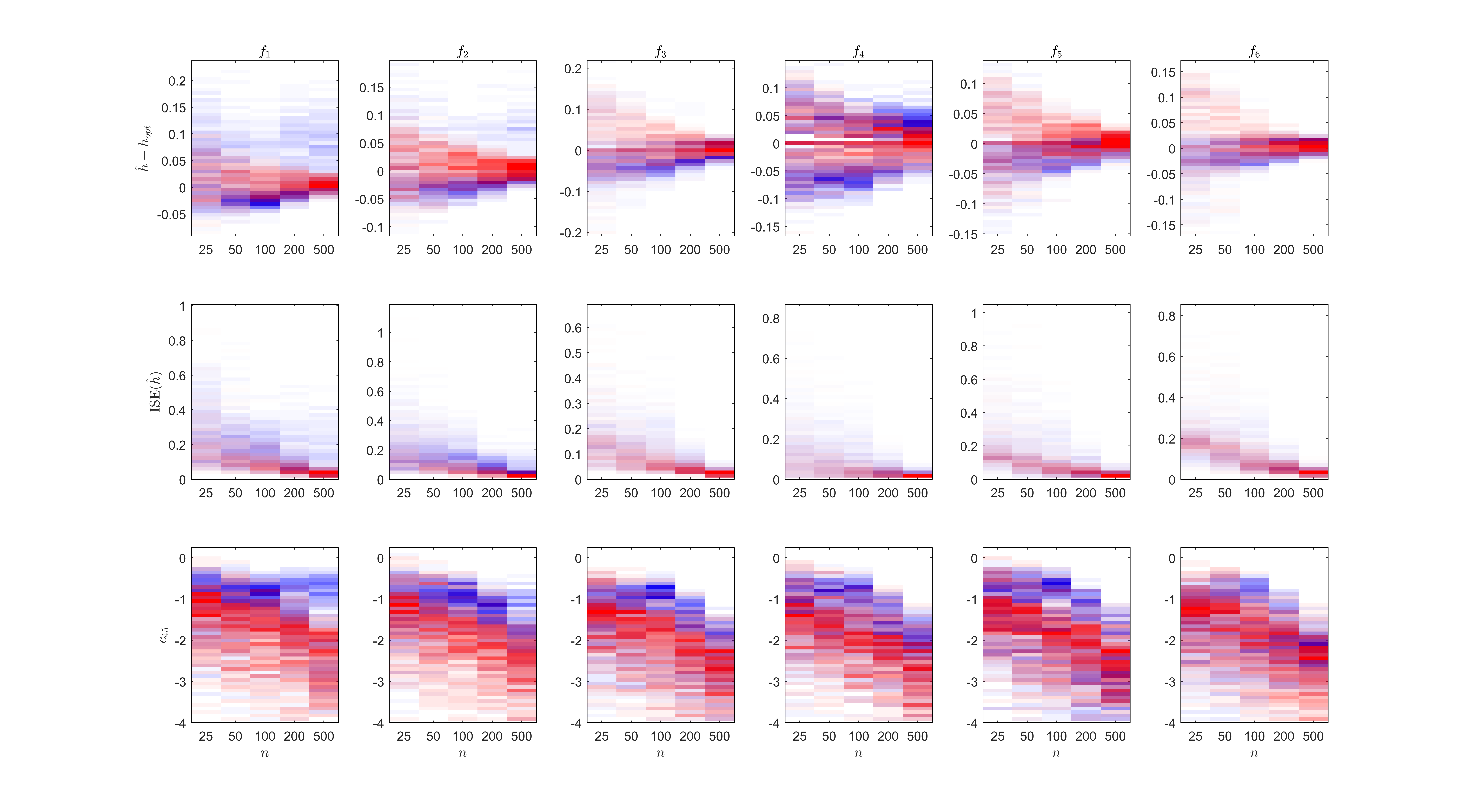}
\caption{ \label{fig:HSSComparisonGauss} Pseudotransparency plots of performance measures for the family $f_{j}$ and varying $n$ using a Gaussian kernel. {\color{red}Data for CV is shown in red}, while {\color{blue}data for TDE is shown in blue}. Each panel represents data on a particular performance measure (top, $\hat h - h_{opt}$; middle, $\text{ISE}(\hat h)$; bottom, $c_{45}$) and $f_j$ (from left to right: $j = 1,\dots,6$). For each value of $n$, the corresponding vertical strip in a panel depicts the histogram (with precisely the same bins for all values of $n$) of a performance measure using a normalized linear transparency scale.}
\end{figure}

\begin{figure}[htbp]
\includegraphics[trim = 35mm 15mm 25mm 10mm, clip, width=180mm,keepaspectratio]{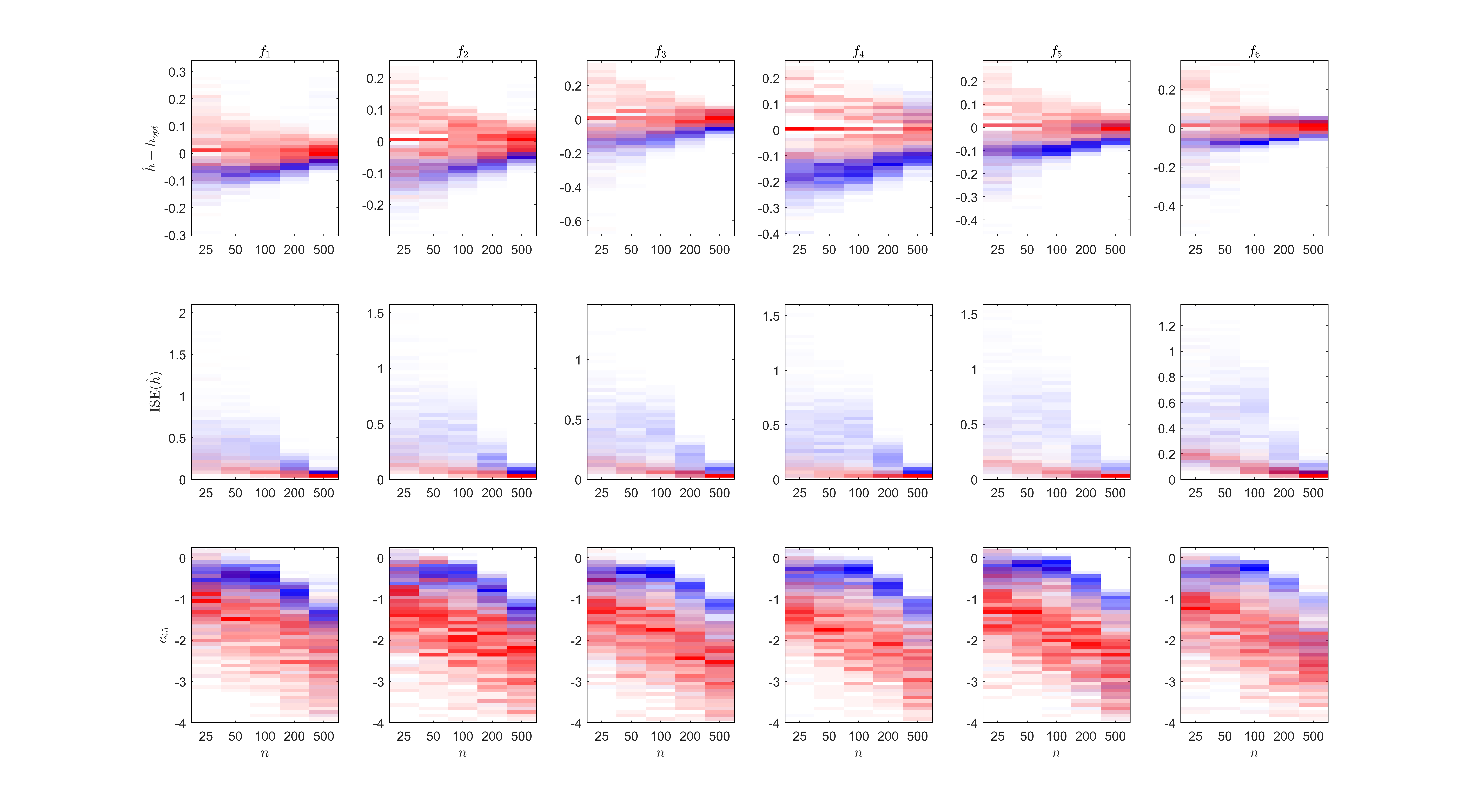}
\caption{ \label{fig:HSSComparisonEpan} As in figure \ref{fig:HSSComparisonGauss}, except that Epanechnikov kernels are used.}
\end{figure}

\begin{figure}[htbp]
\begin{tabular}{c c}
	Gaussian & Epanechnikov \\
	\includegraphics[width=90mm,keepaspectratio]{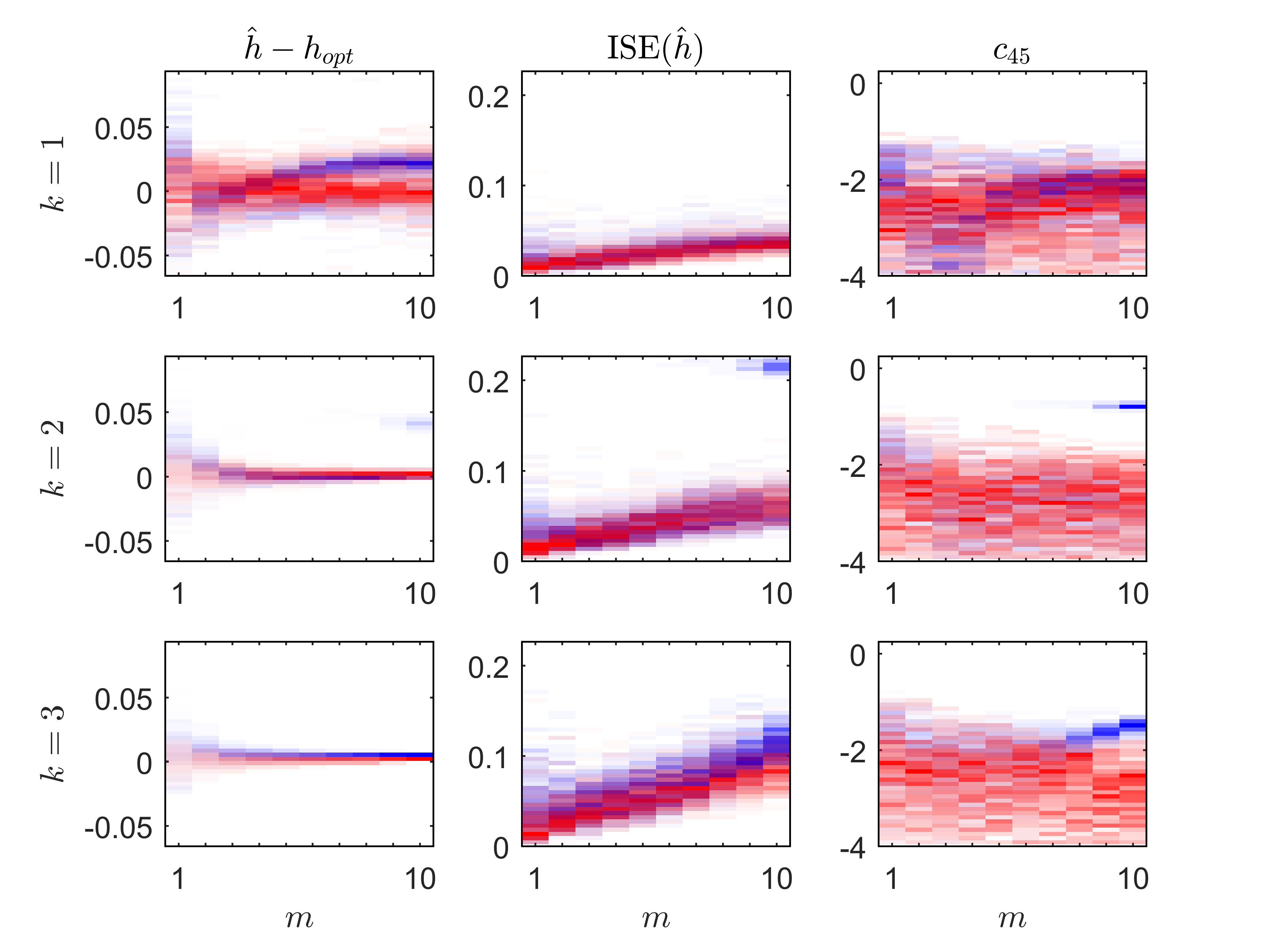}
	&
	\includegraphics[width=90mm,keepaspectratio]{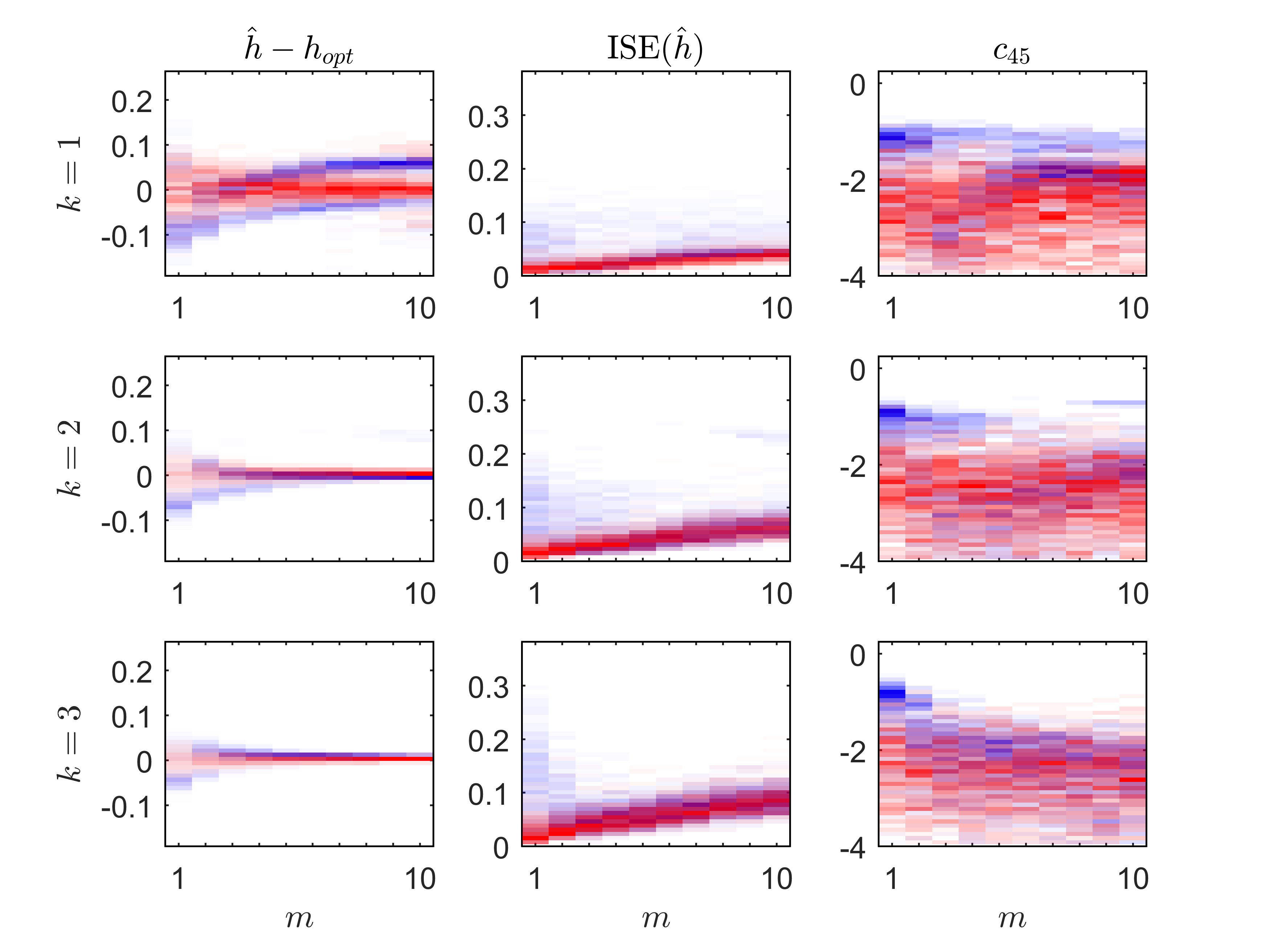}
\end{tabular}
\caption{ \label{fig:fkmEval500Matlab} (Left panels) Pseudotransparency plots of performance measures for the family $f_{km}$ for $n = 500$ using a Gaussian kernel. {\color{red}Data for CV is shown in red}, while {\color{blue}data for TDE is shown in blue}. Each panel represents data on a particular performance measure (left, $\hat h - h_{opt}$; middle, $\text{ISE}(\hat h)$; right, $c_{45}$) and value of $k$ (top, $k = 1$; middle, $k = 2$; bottom, $k = 3$), with $m$ varying from 1 to 10. The vertical axis limits are the same within each column. For each value of $m$, the corresponding vertical strip in a panel depicts the histogram (with precisely the same bins for all values of $m$) of a performance measure using a normalized linear transparency scale. (Right panels) As in the left panels, but with an Epanechnikov kernel.}
\end{figure}

\begin{figure}[htbp]
\begin{tabular}{c c}
	Gaussian & Epanechnikov \\
	\includegraphics[width=90mm,keepaspectratio]{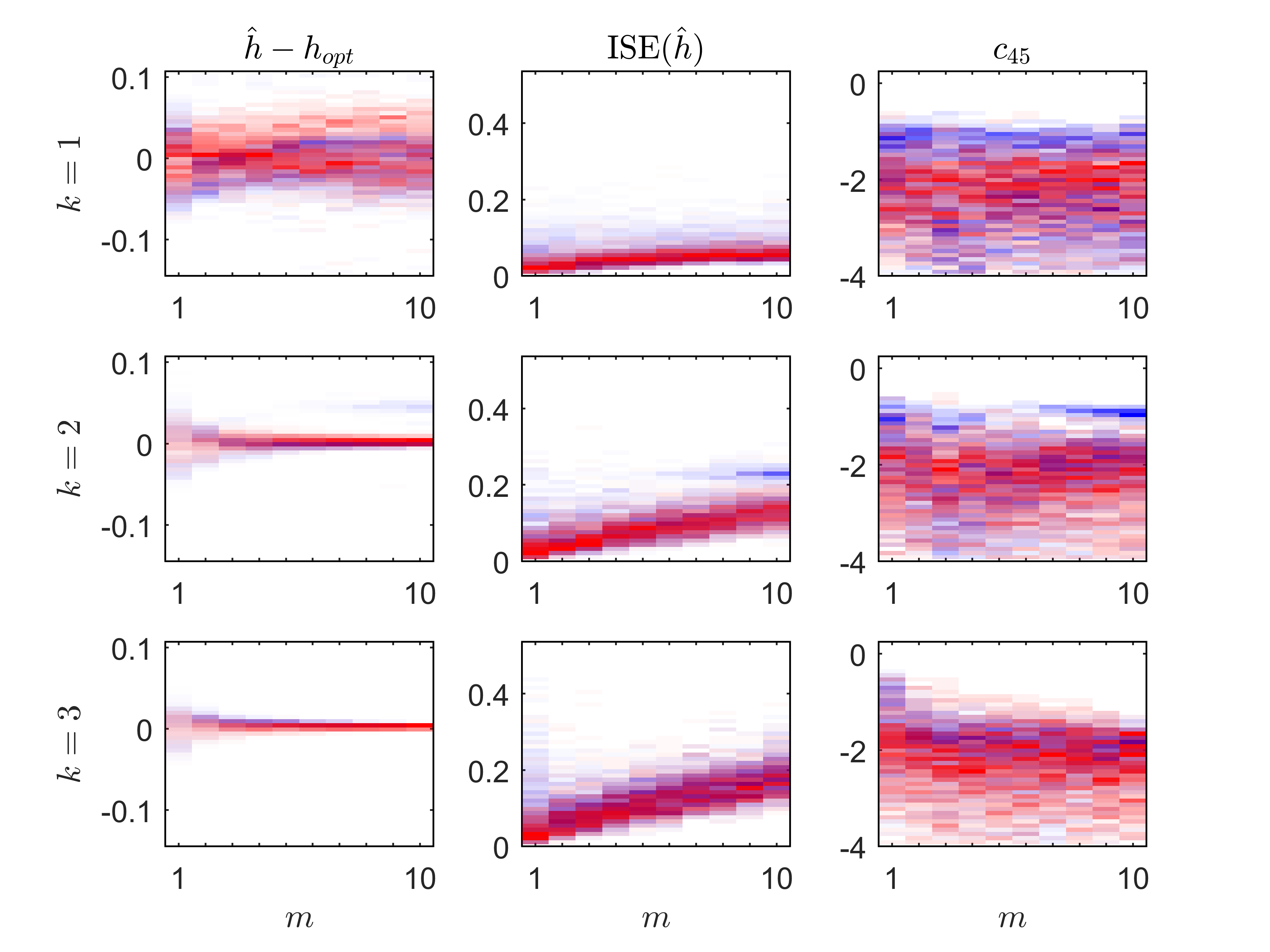}
	&
	\includegraphics[width=90mm,keepaspectratio]{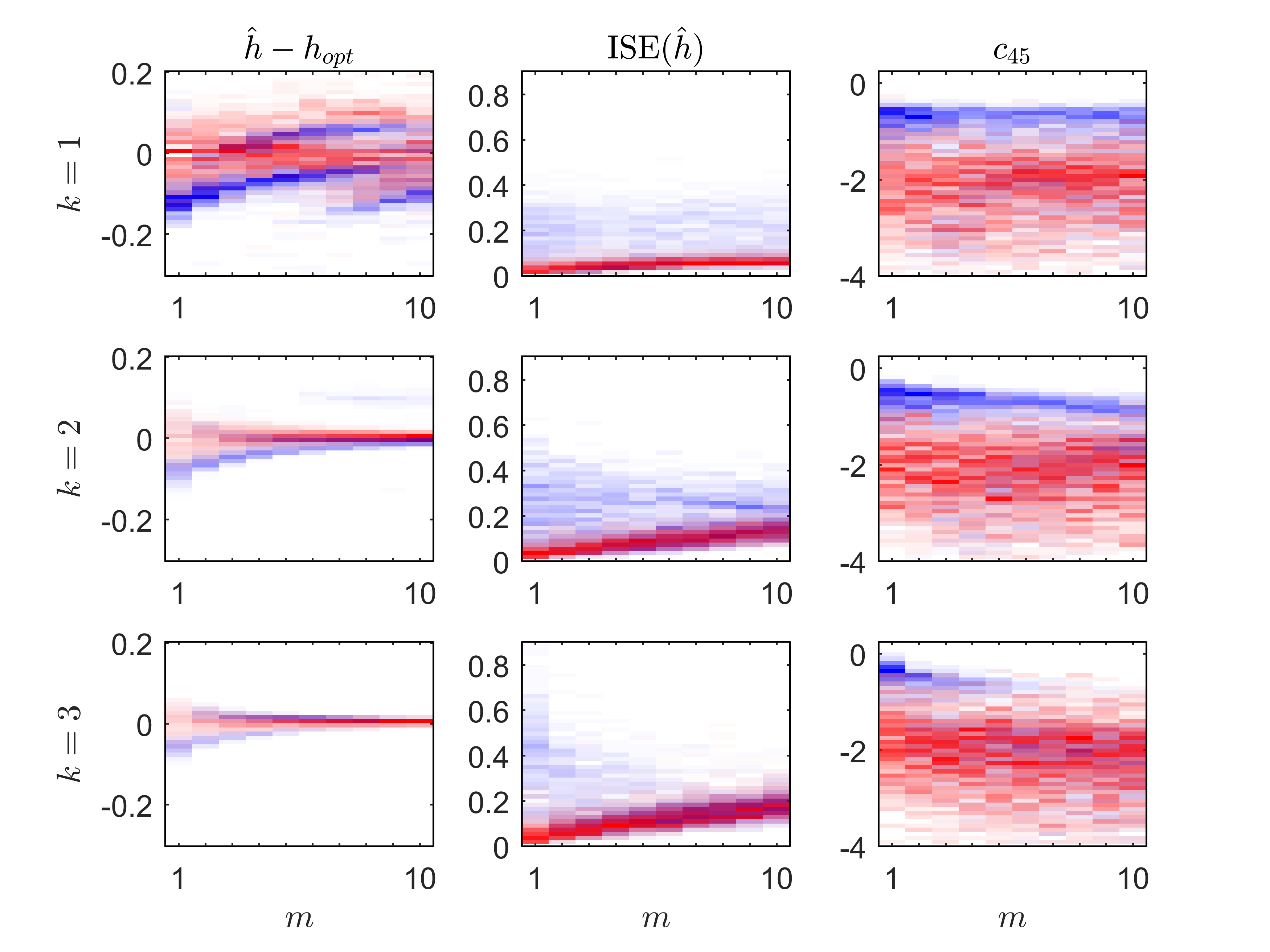}
\end{tabular}
\caption{ \label{fig:fkmEval200Matlab} As in figure \ref{fig:fkmEval500Matlab} for $n = 200$.}
\end{figure}

\begin{figure}[htbp]
\begin{tabular}{c c}
	Unimodal category & Number of local maxima \\
	\includegraphics[width=90mm,keepaspectratio]{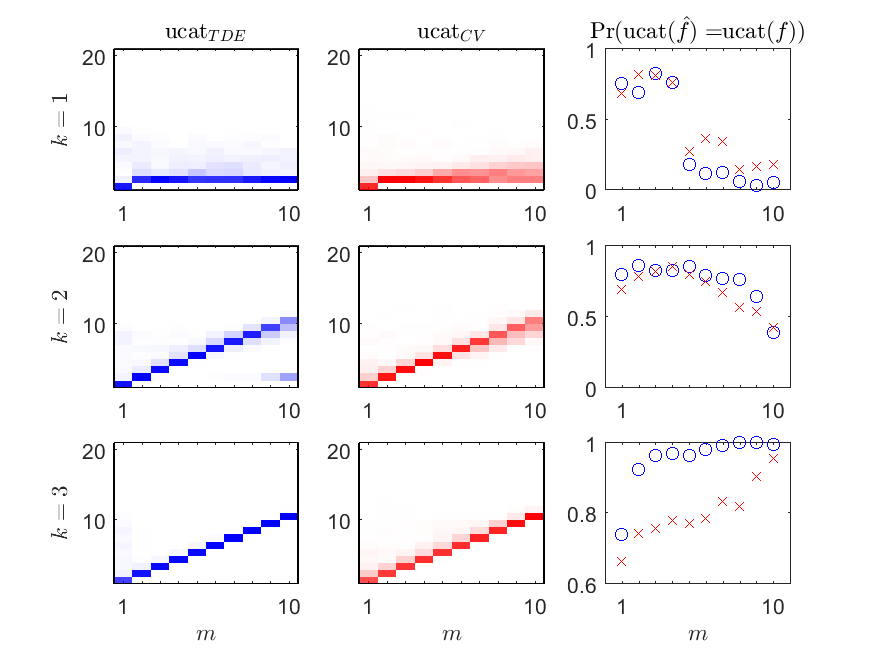}
	&
	\includegraphics[width=90mm,keepaspectratio]{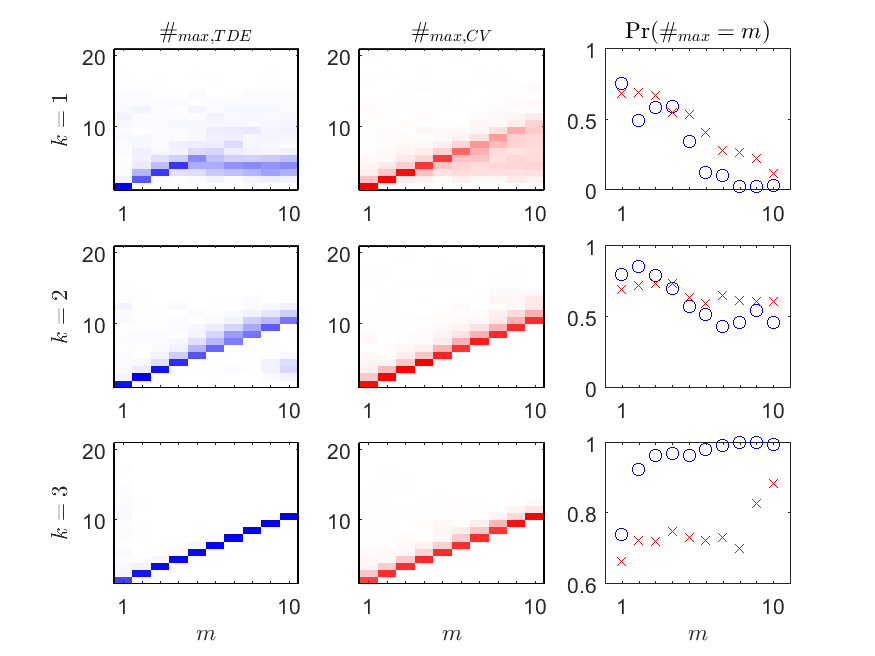}
\end{tabular}
\caption{ \label{fig:fkmEval500MatlabTop} (Left panels) Pseudotransparency plots of performance measures relating to the unimodal category for the family $f_{km}$ for $n = 500$ using a Gaussian kernel. {\color{red}Data for CV is shown in red}, while {\color{blue}data for TDE is shown in blue}. Each panel represents data on a particular performance measure (left, $\text{ucat}$ for TDE alone; middle, $\text{ucat}$ for CV alone; right, the empirical probability that the estimate of $\text{ucat}$ is correct) and value of $k$ (top, $k = 1$; middle, $k = 2$; bottom, $k = 3$), with $m$ varying from 1 to 10. (Right panels) As in the left panels, but for the number of local maxima instead of the unimodal category.}
\end{figure}

\begin{figure}[htbp]
\begin{tabular}{c c}
	Unimodal category & Number of local maxima \\
	\includegraphics[width=90mm,keepaspectratio]{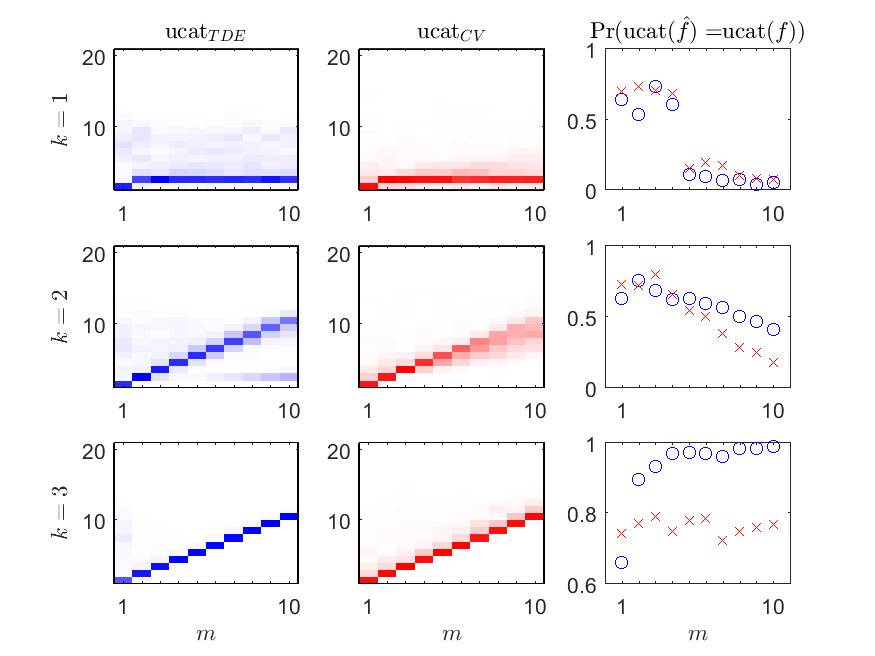}
	&
	\includegraphics[width=90mm,keepaspectratio]{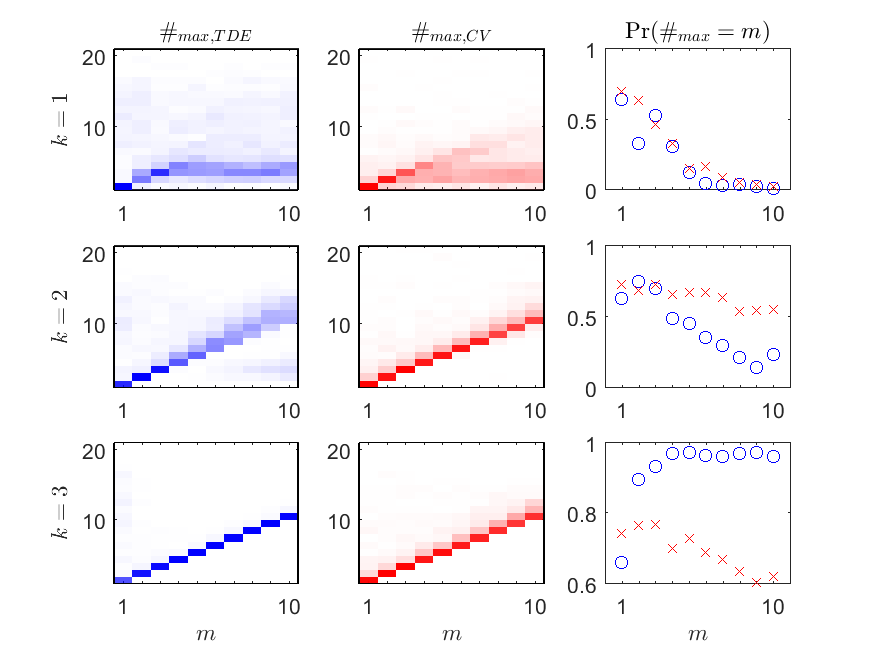}
\end{tabular}
\caption{ \label{fig:fkmEval200MatlabTop} As in figure \ref{fig:fkmEval500Matlab} for $n = 200$.}
\end{figure}

The results of CV and TDE on the family \eqref{eq:f_j} are summarized in figures \ref{fig:HSSComparisonGauss} and \ref{fig:HSSComparisonEpan}. Results of CV and TDE on the family \eqref{eq:fkm} for $n \in \{200,500\}$ are summarized in figures \ref{fig:fkmEval500Matlab}-\ref{fig:fkmEval200MatlabTop}, and results for $n \in \{25,50,100\}$ are summarized in figures \ref{fig:fkmEval100Matlab}-\ref{fig:fkmEval25MatlabTop} in \S \ref{sec:SmallSamples}. 

While TDE underperforms CV on the six densities in \eqref{eq:f_j}, it is still competitive on the three multimodal densities $f_3$, $f_5$, and $f_6$ when using a Gaussian kernel. As we shall see below using \eqref{eq:fkm}, the relative performance of TDE improves with increasing multimodality, to the point that it eventually outperforms CV on qualitative criteria such as the number of local maxima and the unimodal category itself.

The relative performance of TDE is better with a Gaussian kernel than with an Epanechnikov kernel: this pattern persists for the family \eqref{eq:fkm}. Indeed, the relative tendency of TDE to underestimate $\hat h - h_{opt}$ diminishes if a Gaussian versus an Epanechnikov kernel is used. 

As the degree of multimodality increases,  the relative tendency of TDE to underestimate $\hat h - h_{opt}$ eventually disappears altogether, and the performance of TDE is only slightly worse than that of CV. Meanwhile, \cite{HeidenreichSchindlerSperlich} shows that CV outperforms all the other methods considered there with respect to $\hat h - h_{opt}$. Therefore we can conclude that TDE offers very competitive performance for $\hat h - h_{opt}$ (or $c_1$) for highly multimodal densities. 

Since CV is expressly designed to minimize the expected value of $\text{ISE}(\hat h)$ (i.e., $c_2$), it is hardly surprising that TDE does not perform as well in this respect. However, it is remarkable that TDE is still so competitive for the distribution of $\text{ISE}(\hat h)$: indeed, the performance of both techniques is barely distinguishable in many of the multimodal cases. Furthermore, the convergence of TDE with increasing $n$ is clearly comparable to that of CV, which along with its derivatives has the best convergence properties among the techniques in \cite{HeidenreichSchindlerSperlich}. While \cite{HeidenreichSchindlerSperlich} points out that CV gives worse values for $c_3$ than competing methods, the performance gap there is fairly small, and so we can conclude that TDE again offers competitive performance for $\text{ISE}(\hat h)$ (or $c_2$ and $c_3$) for highly multimodal densities.

The only respect in which CV offers a truly qualitative advantage over TDE for highly multimodal densities is $c_{45}$ (or its surrogates $c_4$ and $c_5$). However, there is still considerable overlap in the distributions for TDE and CV, and for multimodal densities and sample sizes of $200$ or more, CV offers nearly the best performance in this respect of all the techniques considered in \cite{HeidenreichSchindlerSperlich}. Therefore we can conclude that TDE offers reasonable though not competitive performance for $c_{45}$ (or $c_4$ and $c_5$) for highly multimodal densities.

The preceding considerations show that TDE is competitive overall with other methods--though still clearly not optimal--for highly multimodal densities (and sample sizes sufficient in principle to resolve the modes) when traditional statistical evaluation criteria are used. However, when \emph{qualitative} criteria such as the number of local maxima and the unimodal category itself are considered, TDE outperforms CV for highly multimodal densities (see figures \ref{fig:fkmEval500MatlabTop} and \ref{fig:fkmEval200MatlabTop}). In practice, such qualitative criteria are generally of paramount importance. For example, precisely estimating the shape of a density is generally less important than determining if it has two or more clearly separable modes. 

Perhaps the most impressive feature of TDE, and one that CV is essentially alone in sharing with it, is the fact that TDE requires no free parameters or assumptions. Indeed, TDE can be used to evaluate its own suitability: for unimodal distributions, it is clearly not an ideal choice--but it is good at detecting this situation in the first place. In fact while for $n = 200$ CV is uniformly better at determining unimodality, for $n = 500$ the situation has reversed, with TDE uniformly better at determining unimodality.

\subsection{\label{sec:Discussion}Discussion}

Having seen \emph{where} and \emph{how} TDE performs well, we now turn to the question of \emph{why} it performs well. 

\begin{figure}[htbp]
\includegraphics[trim = 30mm 170mm 30mm 20mm, clip, width=180mm,keepaspectratio]{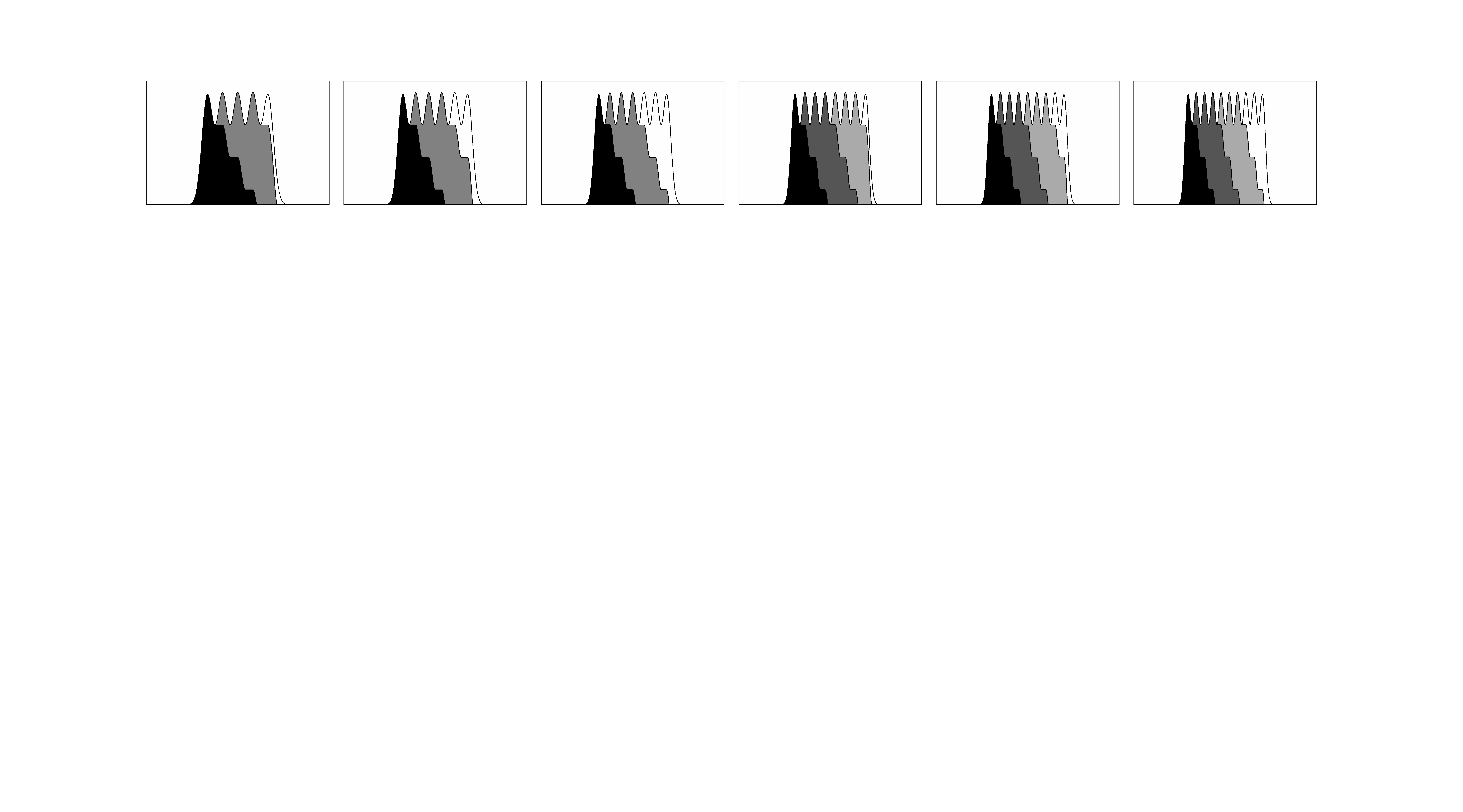}
\caption{ \label{fig:unidec_a} Unimodal decompositions of $f_{km}$ for $k = 1$ and $5 \le m \le 10$. Note that $\text{ucat}(f_{km}) = 3$ for $5 \le m \le 7$ and $\text{ucat}(f_{km}) = 4$ for $8 \le m \le 10$.} 
\end{figure}

\begin{figure}[htbp]
\includegraphics[trim = 30mm 170mm 30mm 20mm, clip, width=180mm,keepaspectratio]{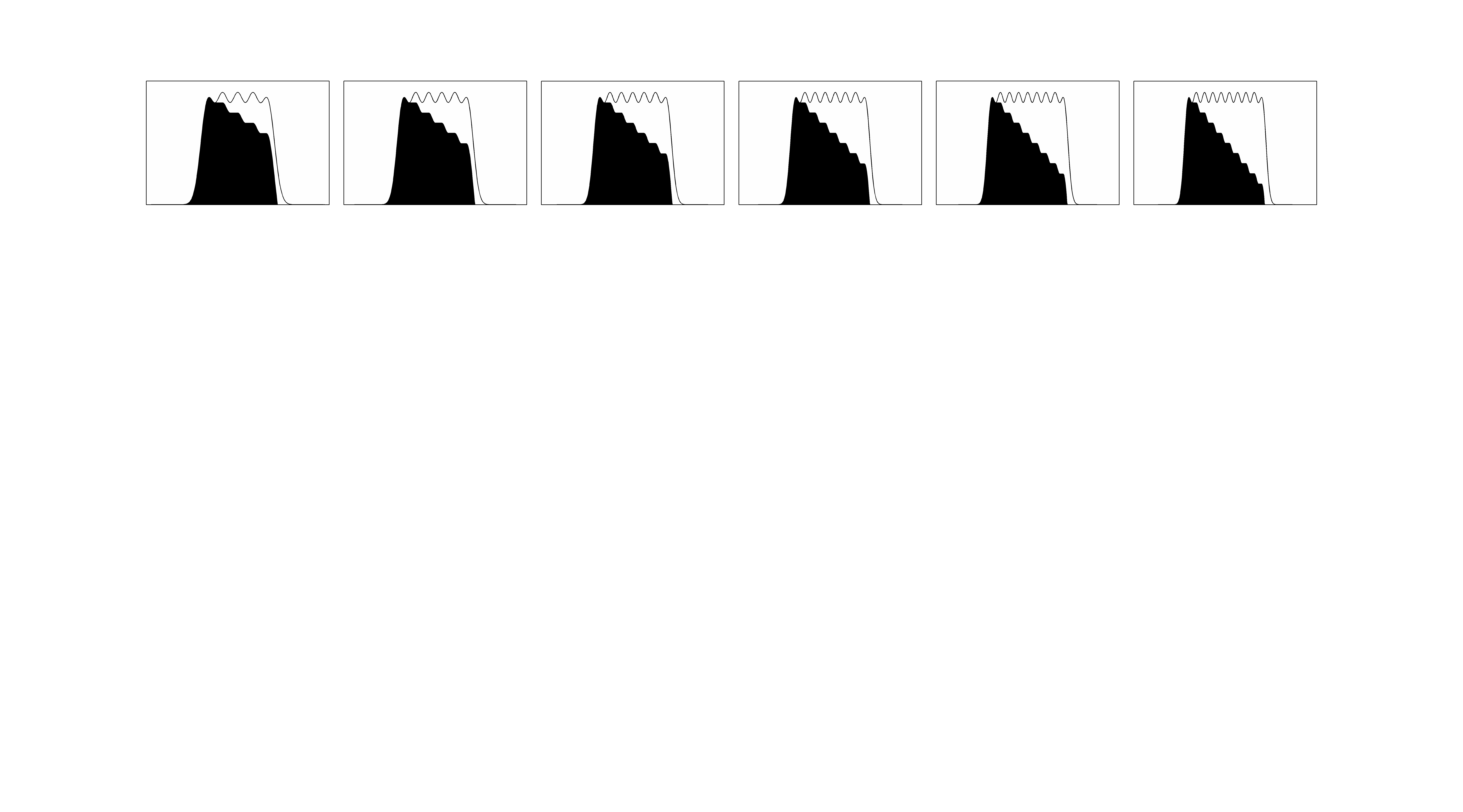}
\caption{ \label{fig:unidec_b} Unimodal decompositions of $f_{km}$ for $k = 0.4$ and $5 \le m \le 10$. Note that $\text{ucat}(f_{km}) = 2$ for $5 \le m \le 10$.} 
\end{figure}

Regarding the family $\{f_{km}\}$, note that if we consider $k$ as a real parameter, for low enough values of $k$ (depending on $m$) the unimodal decomposition will yield less than $m$ (and in the limit, only two) components, whereas for large values of $k$ the unimodal decomposition will closely approximate the underlying Gaussian mixture. (See figures \ref{fig:unidec_a} and \ref{fig:unidec_b}.) TDE works well even for relatively small $k$ since in this case the few-component unimodal decomposition still persists over bandwidths that correctly resolve extrema. This indicates how the nontopological mode-hunting approaches of \cite{Silverman,Minnotte} can be effectively subsumed by TDE.

	%
	%
	%

The basic result of \cite{Silverman} is that for $K$ Gaussian, the number of maxima of $\partial^j \hat f_{h;X} / \partial x^j$ is right-continuous and nonincreasing in $h$ for all $j \ge 0$. One might consider trying to leverage this result by using a Gaussian kernel and taking the number of maxima as a proxy for the unimodal category. However, the number of maxima is less stable than the unimodal category. To see this, again consider figures \ref{fig:unidec_a} and \ref{fig:unidec_b}, in which it is graphically demonstrated that the unimodal category tends to be constant under small perturbations that produce a new mode. This tendency in turn derives from the topological invariance of the unimodal category: for, most small perturbations can be associated with a single component of a unimodal decomposition. Thus the unimodal category gives a stable platform for persistence because it is both an integer and a topological invariant.

\section{\label{sec:Remarks}Remarks}

\subsection{\label{sec:Runtime}Runtime}

An important advantage that TDE offers both in theory and practice relative to CV or more sophisticated KDE techniques is that it is computationally efficient. Both TDE and CV spend most of their time evaluating and summing kernels. If $n_h$ denotes the number of proposed bandwidths and $n_x$ denotes the number of points that a density (not a kernel) is represented (not estimated) on, then TDE requires $n_h n_x n$ kernel evaluations, whereas CV requires $2 n_h n^2$ kernel evaluations. For $n$ large, we can and even should take $n_x \ll n$: e.g., we can adequately represent even a quite complicated density using $n_x = 100$ data points even if $n = 1000$ samples are required to estimate the density in the first place. Meanwhile, the unimodal decompositions themselves contribute only a marginal $O(n_h n_x \langle m \rangle_h)$ operations, where $\langle m \rangle_h \ll n_x$ indicates the average number of modes as the bandwidth varies. 

Thus TDE ought to be (and in practice is) considerably faster than CV for $n$ large. Note also that invoking sophisticated kernel summation techniques such as the fast Gauss transform would confer proportional advantages to both TDE and CV, so the runtime advantage of TDE is essentially fundamental so long as equal care is taken in implementations of the two techniques.

\subsection{\label{sec:Modifications}Modifications}

In \cite{Minnotte} a ``mode tree'' is used to visually evaluate an adaptive density estimate in which potential modes are estimated with independent (but fixed) bandwidths. This construction anticipates TDE in some ways and generalizes it in others. However, these generalizations can be reproduced in the TDE framework: for instance, it is possible to use the result of TDE as an input to an explicit unimodal decomposition, then ``pull back'' the decomposition to sample data and finally apply CV or another appropriate method of the type considered in \cite{HeidenreichSchindlerSperlich} independently to each mode. It seems likely that such a technique would offer further improvements in performance for moderately multimodal densities, while still requiring no free configuration choices or parameters. However, it would probably be more desirable to seek a unimodal decomposition minimizing the Jensen-Shannon divergence, and while heuristics can certainly guide a local search to reduce the Jensen-Shannon divergence, it is not \emph{a priori} obvious that a global minimum can be readily obtained.

There is also an alternative to \eqref{eq:TDE} that is worth considering: let $\hat x_j(h)$ for $1 \le j \le \text{ucat}(\hat f_h)$ denote the maxima of the components of a unimodal decomposition of $\hat f_h$. Write 
\begin{equation}
\bar {\hat x}_j := \frac{1}{\mu(u_X^{-1}(\hat m))} \int_{u_X^{-1}(\hat m)} \hat x_j(h) \ dh \nonumber
\end{equation}
and take
\begin{equation}
\label{eq:TDEalternative}
\hat h := \arg \min_{h \in u_X^{-1}(\hat m)} \| \hat x(h) - \bar {\hat x} \|. 
\end{equation}
That is, take $\hat h$ to be the bandwidth that produces modes whose loci of maxima are the most stable. This (or something like it) also seems likely to improve performance.

\subsection{\label{sec:HigherDimensions}Extension to higher dimensions}

Surprisingly, learning Gaussian mixture models (GMMs) becomes easier as the spatial dimension increases \cite{AndersonEtAl}. However, this result requires knowledge of the number of components of a GMM, which is precisely the sort of datum that high-dimensional TDE would provide. This shows that an extension to all $d > 1$ is very desirable.

However, it is not even clear how to extend TDE to $d = 2$. While in this case a result of \cite{BaryshnikovGhrist} shows that $\text{ucat}(f)$ is a function of the combinatorial type of the Reeb graph of $f$ labeled by critical values (i.e., the contour tree \cite{CarrSnoeyinkAxen}), the critical values of $\hat f_h$ will vary with $h$ and an explicit algorithm for computing $\text{ucat}(\hat f_h)$ is still lacking.

\acknowledgements

The author is grateful to Jeong-O Jeong for the detection and remediation of a subtle and annoying bug in the author's MATLAB implementation of the unimodal decomposition; to Facundo M\'emoli for an illuminating conversation that helped translate the ideas in this paper from the discrete setting where they were initially developed and deployed; and to Robert Ghrist for hosting the author on a visit, the preparations for which led to several important corrections and observations in the present version of this paper.

\appendix

\section{\label{sec:MATLABScripts}MATLAB scripts}

NB. The statistics toolbox is required for scripts that generate evaluation data; however, it is not required for most of the code in \S \ref{sec:MATLABFunctions}.

\subsection{\label{sec:TDEIndividualPlotsScript}TDEIndividualPlotsScript.m}

\footnotesize
\begin{verbatim}
% Script for plotting the densities $f_j$ ($1 \le j \le 6$) and $f_{km}$
% ($1 \le k \le 3$, $1 \le m \le 10$), generating individual sample sets
% from each, performing CV and TDE, and visualizing the results.
%
% NB. The statistics toolbox is required for this script.

rng('default');

%% Script configuration
% Linear space
x_0 = -1;
x_1 = 2;
n_x = 500;  % number of evaluation points for PDFs
x = linspace(x_0,x_1,n_x);
% Number of samples
n = 200;
% Kernel
kernelflag = 1;
if kernelflag > 0
    kernelargstr = 'kernel';
    titlestr = 'Gaussian';
elseif kernelflag < 0
    kernelargstr = 'epanechnikov';
    titlestr = 'Epanechnikov';
else
    error('bad kernel flag');
end

%% Generate samples and PDFs
[X,f] = tdepdfsuite(n_x,n);

%% Plot (used to generate f_j.png)
figure;
ha = tight_subplot(5,6,[.01 .01],[.1 .1],[.1 .1]);
for j = 1:6
    axes(ha(j));    
    plot(x,f(j,:),'k');
    xlim([-.5,1.5])
    ymax = 1.1*max(f(j,:));
    ylim([0,ymax]);
    set(ha(j),'XTick',[],'YTick',[]);
end
for j = 7:30, axes(ha(j)); axis off; end

%% Plot (used to generate f_km.png)
figure;
Mvert = 5;
Mhorz = 6;
ha = tight_subplot(Mvert,Mhorz,[.01 .01],[.1 .1],[.1 .1]);
for k = 1:3
    for m = 1:Mhorz
        j = (k-1)*Mhorz+m;
        km = 6+(k-1)*10+m;
        axes(ha(j));    
        plot(x,f(km,:),'k');
        xlim([-.5,1.5]);
        ymax = 1.1*max(f(km,:));
        ylim([0,ymax]);
        set(ha(j),'XTick',[],'YTick',[]);
    end
end

%% Density estimates and plots 
for j = 1:size(X,1)
    %% Optimal kernel density estimate w/ knowledge of PDF
    opt = optimalkernelbandwidth(x,f(j,:),n,kernelflag);
    pd_opt = fitdist(X(j,:)',kernelargstr,'BandWidth',opt.h);
    kde_opt = pdf(pd_opt,x);
    %% Numerically ISE minimizing bandwidth
    temp = vbise(x,f(j,:),X(j,:),kernelflag);
    ind = find(temp.ISE==min(temp.ISE),1,'first');
    h_opt = temp.h(ind);
    pd_numopt = fitdist(X(j,:)',kernelargstr,'BandWidth',h_opt);
    kde_numopt = pdf(pd_numopt,x);
    %% Cross-validation KDE
    pd = fitdist(X(j,:)',kernelargstr);
    kde = pdf(pd,x);
    %% Topological KDE
    tde = tde1d(X(j,:),kernelflag);
    f_tde = tde.y/(sum(tde.y)*mean(diff(tde.x)));
    %% Plot
    figure;
    plot(x,f(j,:),'k',...
        x,kde_opt,'b',...
        x,kde_numopt,'c',...
        x,kde,'g',...
        tde.x,f_tde,'r');
    xlim([-.5,1.5])
    set(gca,'XTick',[],'YTick',[]);
    title([titlestr,' kernel density estimates of PDF'],...
        'Interpreter','latex');
    legend({'PDF',...
        ['$\hat h_0 = $',num2str(opt.h,'%0.4f')],...
        ['$h_{opt} = $',num2str(h_opt,'%0.4f')],...
        ['$\hat h_{CV} = $',num2str(pd.BandWidth,'%0.4f')],...
        ['$\hat h_{top} = $',num2str(tde.h,'%0.4f')]},...
        'Interpreter','latex');
end
\end{verbatim}
\normalsize
\vskip\medskipamount 
\leaders\vrule width \textwidth\vskip0.4pt 
\vskip\medskipamount 
\nointerlineskip

\subsection{\label{sec:TDEPerformanceDataScript}TDEPerformanceDataScript.m}

\footnotesize
\begin{verbatim}
% Script for generating performance data for evaluating CV and TDE on the
% densities $f_j$ ($1 \le j \le 6$) and $f_{km}$ ($1 \le k \le 3$, $1 \le m
% \le 10$).
%
% NB. The statistics toolbox is required for this script.

rng('default');

%% Preliminaries and script configuration
% Linear space
x_0 = -1;
x_1 = 2;
n_x = 500;
x = linspace(x_0,x_1,n_x);
% Number of simulation runs
N = 250;
% Number of samples
n = [25,50,100,200,500];	% we omit n = 1000
% Kernel
kernelflag = 1;
if kernelflag > 0
    kernelargstr = 'kernel';
    titlestr = 'Gaussian';
elseif kernelflag < 0
    kernelargstr = 'epanechnikov';
    titlestr = 'Epanechnikov';
else
    error('bad kernel flag');
end

%% Main loop
for aa = 1:numel(n)
    %% Loop over simulation runs
    for ii = 1:N
        disp([aa,ii]);
        %% Generate samples and PDFs
        [X,f] = tdepdfsuite(n_x,n(aa));
        %% Loop over PDFs
        for j = 1:size(X,1) 
            %% Optimal kernel density estimate w/ knowledge of PDF
            opt = optimalkernelbandwidth(x,f(j,:),n(aa),kernelflag);
            if kernelflag > 0
                pd_opt = fitdist(X(j,:)',kernelargstr,'BandWidth',opt.h);
            elseif kernelflag < 0
                pd_opt = fitdist(X(j,:)','Kernel','Kernel',kernelargstr,...
                    'BandWidth',opt.h);
            end
            kde_opt = pdf(pd_opt,x);
            %% Numerically ISE minimizing bandwidth
            temp = vbise(x,f(j,:),X(j,:),kernelflag);
            ind = find(temp.ISE==min(temp.ISE),1,'first');
            h_opt = temp.h(ind);
            if kernelflag > 0
                pd_numopt = fitdist(X(j,:)',kernelargstr,'BandWidth',h_opt);
            elseif kernelflag < 0
                pd_numopt = fitdist(X(j,:)','Kernel','Kernel',...
                    kernelargstr,'BandWidth',h_opt);
            end
            kde_numopt = pdf(pd_numopt,x);
            %% Cross-validation KDE
            cv = cv1d(X(j,:)',kernelflag);
            if kernelflag > 0
                % % For MATLAB estimator
                % pd = fitdist(X(j,:)',kernelargstr);   
                pd = fitdist(X(j,:)',kernelargstr,'BandWidth',cv.h);
            elseif kernelflag < 0
                % % For MATLAB estimator
                % pd = fitdist(X(j,:)','Kernel','Kernel',kernelargstr);
                pd = fitdist(X(j,:)','Kernel','Kernel',kernelargstr,'BandWidth',cv.h);
            end
            kde = pdf(pd,x);
            h_hat = pd.BandWidth;
            pre_c1_CV(ii,j,aa) = h_hat-h_opt;
            pre_c23_CV(ii,j,aa) = ise(x,f(j,:),X(j,:),h_hat,kernelflag);
            pre_c45_CV(ii,j,aa) = ...
                pre_c23_CV(ii,j,aa)-ise(x,f(j,:),X(j,:),h_opt,kernelflag);
            %% Topological KDE
            tde = tde1d(X(j,:),kernelflag);
            f_tde = tde.y/(sum(tde.y)*mean(diff(tde.x)));
            h_hat = tde.h;
            pre_c1_top(ii,j,aa) = h_hat-h_opt;
            pre_c23_top(ii,j,aa) = ise(x,f(j,:),X(j,:),h_hat,kernelflag);
            pre_c45_top(ii,j,aa) = ...
                pre_c23_top(ii,j,aa)-ise(x,f(j,:),X(j,:),h_opt,kernelflag);
            %% Unimodal category and number of local maxima
            ucat_CV(ii,j,aa) = nnz(sum(unidec(kde,0),2)>sqrt(eps));
            lmax_CV(ii,j,aa) = nnz(diff(sign(diff([0,kde,0])))<0);
            ucat_top(ii,j,aa) = tde.mfuc;
            lmax_top(ii,j,aa) = nnz(diff(sign(diff([0,f_tde,0])))<0);
        end  
    end
end
\end{verbatim}
\normalsize
\vskip\medskipamount 
\leaders\vrule width \textwidth\vskip0.4pt 
\vskip\medskipamount 
\nointerlineskip

\subsection{\label{sec:TDEEnsemblePlotsScriptFj}TDEEnsemblePlotsScriptFj.m}

\footnotesize
\begin{verbatim}
% Script for plotting performance data for evaluating CV and TDE on the
% densities $f_j$ ($1 \le j \le 6$).

figure;

%% Normalization macro
normmacro = ['A = A./repmat(sum(A,1),[size(A,1),1]);',...
    'B = B./repmat(sum(B,1),[size(B,1),1]);'];

%% Plot macro
% "box on" gives inadequate results but retained for tickmarks; manually
% set axis and put in a box since "box on" doesn't give adequate results
plotmacro = ['layertrans2(0:5,[L,Inf],A,B);box on;',... 
    'set(gca,''XTick'',1:5,''XTickLabel'',',...
    '{''25'',''50'',''100'',''200'',''500''});',...
    'hold on;axis([0.5,5.5,lo,hi]);ax = axis;',...
    'line([ax(1),ax(1)]+1e-6,[ax(3),ax(4)],''Color'',''k'');',...
    'line([ax(2),ax(2)]-1e-6,[ax(3),ax(4)],''Color'',''k'');',...
    'line([ax(1),ax(2)],[ax(3),ax(3)]+1e-6,''Color'',''k'');',...
    'line([ax(1),ax(2)],[ax(4),ax(4)]-1e-6,''Color'',''k'');'];

%% c1
for j = 1:6
    % Normalized histogram
    lo = min([min(min(pre_c1_top(:,j,:))),min(min(pre_c1_CV(:,j,:)))]);
    hi = max([max(max(pre_c1_top(:,j,:))),max(max(pre_c1_CV(:,j,:)))]);
    L = linspace(lo,hi,50);
    for k = 1:5 % n = [25,50,100,200,500]
        A(:,k) = histc(squeeze(pre_c1_top(:,j,k)),L,1); 
        B(:,k) = histc(squeeze(pre_c1_CV(:,j,k)),L,1); 
    end
    eval(normmacro);
    % Plot
    subplot(3,6,j);  
    eval(plotmacro);
    title(['$f_',num2str(j),'$'],'Interpreter','latex');
end
subplot(3,6,1);
ylabel('$\hat h - h_{opt}$','Interpreter','latex');

%% c2, c3
for j = 1:6
    % Normalized histogram
    lo = 0;
    hi = max([max(max(pre_c23_top(:,j,:))),max(max(pre_c23_CV(:,j,:)))]);
    L = linspace(lo,hi,50);
    for k = 1:5 % n = [25,50,100,200,500]
        A(:,k) = histc(squeeze(pre_c23_top(:,j,k)),L,1); 
        B(:,k) = histc(squeeze(pre_c23_CV(:,j,k)),L,1); 
    end
    eval(normmacro);
    % Plot
    subplot(3,6,j+6);  
    eval(plotmacro);
end
subplot(3,6,7);
ylabel('ISE$(\hat h)$','Interpreter','latex');

%% c4, c5
for j = 1:6
    % Form histograms (note logspace). NB. The only difference between L1
    % and L2 versions is rescaling    
    lo = -4;
    hi = 0.25;
    L = linspace(lo,hi,50);
    % abs turns out to not have any effect, but included for correctness
    for k = 1:5 % n = [25,50,100,200,500]
        A(:,k) = histc(squeeze(log10(abs(pre_c45_top(:,j,k)))),L,1); 
        B(:,k) = histc(squeeze(log10(abs(pre_c45_CV(:,j,k)))),L,1); 
    end
    eval(normmacro);    
    % Plot
    subplot(3,6,j+12);  
    eval(plotmacro);
    xlabel('$n$','Interpreter','latex');
end
subplot(3,6,13);
ylabel('$c_{45}$','Interpreter','latex');
\end{verbatim}
\normalsize
\vskip\medskipamount 
\leaders\vrule width \textwidth\vskip0.4pt 
\vskip\medskipamount 
\nointerlineskip

\subsection{\label{sec:TDEEnsemblePlotsScriptFkm}TDEEnsemblePlotsScriptFkm.m}

\footnotesize
\begin{verbatim}
% Script for plotting performance data for evaluating CV and TDE on the
% densities $f_{km}$ ($1 \le k \le 3$, $1 \le m \le 10$).

%% Normalization macro
normmacro = ['A = A./repmat(sum(A,1),[size(A,1),1]);',...
    'B = B./repmat(sum(B,1),[size(B,1),1]);'];

%% Plot macro
% "box on" gives inadequate results but retained for tickmarks; manually
% set axis and put in a box since "box on" doesn't give adequate results
plotmacro = ['layertrans2(0:10,[L,Inf],A(:,ind),B(:,ind));box on;',... 
    'set(gca,''XTick'',1:10,''XTickLabel'',',...
    '{''1'','''','''','''','''','''','''','''','''',''10''});',...
    'hold on;axis([0.5,10.5,lo,hi]);ax = axis;',...
    'line([ax(1),ax(1)]+1e-6,[ax(3),ax(4)],''Color'',''k'');',...
    'line([ax(2),ax(2)]-1e-6,[ax(3),ax(4)],''Color'',''k'');',...
    'line([ax(1),ax(2)],[ax(3),ax(3)]+1e-6,''Color'',''k'');',...
    'line([ax(1),ax(2)],[ax(4),ax(4)]-1e-6,''Color'',''k'');'];

%% Loop over values of n
for j = 1:5 % n = [25,50,100,200,500]
    figure;

    %% c1
    % Form histograms
    lo = min([min(min(pre_c1_top(:,:,j))),min(min(pre_c1_CV(:,:,j)))]);
    hi = max([max(max(pre_c1_top(:,:,j))),max(max(pre_c1_CV(:,:,j)))]);
    L = linspace(lo,hi,50);
    A = histc(pre_c1_top(:,:,j),L,1); 
    B = histc(pre_c1_CV(:,:,j),L,1); 
    eval(normmacro);
    % Transparency plots
    subplot(3,3,1); ind = 7:16; eval(plotmacro);
    ylabel('$k=1$','Interpreter','latex');
    title('$\hat h - h_{opt}$','Interpreter','latex');
    subplot(3,3,4); ind = 17:26; eval(plotmacro);
    ylabel('$k=2$','Interpreter','latex');
    subplot(3,3,7); ind = 27:36; eval(plotmacro);
    xlabel('$m$','Interpreter','latex');
    ylabel('$k=3$','Interpreter','latex');

    %% c2, c3
    % Form histograms
    lo = 0;
    hi = max([max(max(pre_c23_top(:,:,j))),max(max(pre_c23_CV(:,:,j)))]);
    L = linspace(lo,hi,50);
    A = histc(pre_c23_top(:,:,j),L,1); 
    B = histc(pre_c23_CV(:,:,j),L,1); 
    eval(normmacro);
    % Transparency plots
    subplot(3,3,2); ind = 7:16; eval(plotmacro);
    title('ISE$(\hat h)$','Interpreter','latex');
    subplot(3,3,5); ind = 17:26; eval(plotmacro);
    subplot(3,3,8); ind = 27:36; eval(plotmacro);
    xlabel('$m$','Interpreter','latex');

    %% c4, c5
    % Form histograms (note logspace). NB. The only difference between L1
    % and L2 versions is rescaling
    % The following is roughly equivalent in practice to 
    %   temp = cat(3,pre_c45_top(:,:,j),pre_c45_CV(:,:,j));
    %   qlog = .05; % quantile margin for log cutoff
    %   lo = log10(quantile(temp(:),qlog));
    %   hi = max([0,log10(max(temp(:)))]);
    % but has the distinct advantage of being constant
    lo = -4;    
    hi = 0.25;
    L = linspace(lo,hi,50);
    % abs turns out to not have any effect, but included for correctness
    A = histc(log10(abs(pre_c45_top(:,:,j))),L,1); 
    B = histc(log10(abs(pre_c45_CV(:,:,j))),L,1); 
    eval(normmacro);
    % Transparency plots
    subplot(3,3,3); ind = 7:16; eval(plotmacro);
	title('$c_{45}$','Interpreter','latex');
    subplot(3,3,6); ind = 17:26; eval(plotmacro);
    subplot(3,3,9); ind = 27:36; eval(plotmacro);
    xlabel('$m$','Interpreter','latex');
    
    %% Figure output
    print('-dpng',[titlestr,'_',num2str(j),'.png']);
end

%% ucat
for i = 1:size(f,1)
    u = unidec(f(i,:),0); 
    ucat(i) = size(u,1); 
end
for j = 1:5
    figure;
    % Form histograms
    lo = 1;
    hi = 21;
    L = 0.5:1:20.5;
    A = histc(ucat_top(:,:,j),L,1); 
    B = zeros(size(A)); B(end,:) = 1; 
    eval(normmacro);    
    % Transparency plots
    subplot(3,3,1); ind = 7:16; eval(plotmacro);
    ylabel('$k=1$','Interpreter','latex');
	title('ucat$_{TDE}$','Interpreter','latex');
    subplot(3,3,4); ind = 17:26; eval(plotmacro);
    ylabel('$k=2$','Interpreter','latex');
    subplot(3,3,7); ind = 27:36; eval(plotmacro);    
    xlabel('$m$','Interpreter','latex');
    ylabel('$k=3$','Interpreter','latex');
    A = B;
    B = histc(ucat_CV(:,:,j),L,1); 
    eval(normmacro);    
    subplot(3,3,2); ind = 7:16; eval(plotmacro);
	title('ucat$_{CV}$','Interpreter','latex');
    subplot(3,3,5); ind = 17:26; eval(plotmacro);
    subplot(3,3,8); ind = 27:36; eval(plotmacro);    
    xlabel('$m$','Interpreter','latex');
    % Diagonal plots
    A = histc(ucat_top(:,:,j),L,1); 
    B = histc(ucat_CV(:,:,j),L,1);
    subplot(3,3,3); ind = 7:16; 
    for i = 1:numel(ind)
        tempA(i) = A(ucat(ind(i)),ind(i))/N;
        tempB(i) = B(ucat(ind(i)),ind(i))/N;
    end
    plot(1:10,tempA,'bo',1:10,tempB,'rx'); xlim([0,11]);
    set(gca,'XTick',1:10,'XTickLabel',{'1','','','','','','','','','10'});
	title('Pr$($ucat$(\hat f)=$ucat$(f))$','Interpreter','latex');
    subplot(3,3,6); ind = 17:26; 
    for i = 1:numel(ind)
        tempA(i) = A(ucat(ind(i)),ind(i))/N;
        tempB(i) = B(ucat(ind(i)),ind(i))/N;
    end
    plot(1:10,tempA,'bo',1:10,tempB,'rx'); xlim([0,11]);
    set(gca,'XTick',1:10,'XTickLabel',{'1','','','','','','','','','10'});
    subplot(3,3,9); ind = 27:36; 
    for i = 1:numel(ind)
        tempA(i) = A(ucat(ind(i)),ind(i))/N;
        tempB(i) = B(ucat(ind(i)),ind(i))/N;
    end
    plot(1:10,tempA,'bo',1:10,tempB,'rx'); xlim([0,11]);
    set(gca,'XTick',1:10,'XTickLabel',{'1','','','','','','','','','10'});
    xlabel('$m$','Interpreter','latex');
    %% Figure output
    print('-dpng',['UCAT',titlestr,'_',num2str(j),'.png']);
end

%% lmax
for j = 1:5
    figure;
    % Form histograms
    lo = 1;
    hi = 21;
    L = 0.5:1:20.5;
    A = histc(lmax_top(:,:,j),L,1); 
    B = zeros(size(A)); B(end,:) = 1; 
    eval(normmacro);    
    % Transparency plots
    subplot(3,3,1); ind = 7:16; eval(plotmacro);
    ylabel('$k=1$','Interpreter','latex');
	title('$\#_{max,TDE}$','Interpreter','latex');
    subplot(3,3,4); ind = 17:26; eval(plotmacro);
    ylabel('$k=2$','Interpreter','latex');
    subplot(3,3,7); ind = 27:36; eval(plotmacro);    
    xlabel('$m$','Interpreter','latex');
    ylabel('$k=3$','Interpreter','latex');
    A = B;
    B = histc(lmax_CV(:,:,j),L,1); 
    eval(normmacro);    
    subplot(3,3,2); ind = 7:16; eval(plotmacro);
	title('$\#_{max,CV}$','Interpreter','latex');
    subplot(3,3,5); ind = 17:26; eval(plotmacro);
    subplot(3,3,8); ind = 27:36; eval(plotmacro);    
    xlabel('$m$','Interpreter','latex');
    % Diagonal plots
    A = histc(lmax_top(:,:,j),L,1); 
    B = histc(lmax_CV(:,:,j),L,1);
    subplot(3,3,3); ind = 7:16; 
    tempA = diag(A(:,ind))/N; tempB = diag(B(:,ind))/N;
    plot(1:10,tempA,'bo',1:10,tempB,'rx'); xlim([0,11]);
    set(gca,'XTick',1:10,'XTickLabel',{'1','','','','','','','','','10'});
	title('Pr$(\#_{max} = m)$','Interpreter','latex');
    subplot(3,3,6); ind = 17:26; 
    tempA = diag(A(:,ind))/N; tempB = diag(B(:,ind))/N;
    plot(1:10,tempA,'bo',1:10,tempB,'rx'); xlim([0,11]);
    set(gca,'XTick',1:10,'XTickLabel',{'1','','','','','','','','','10'});
    subplot(3,3,9); ind = 27:36; 
    tempA = diag(A(:,ind))/N; tempB = diag(B(:,ind))/N;
    plot(1:10,tempA,'bo',1:10,tempB,'rx'); xlim([0,11]);
    set(gca,'XTick',1:10,'XTickLabel',{'1','','','','','','','','','10'});
    xlabel('$m$','Interpreter','latex');
    %% Figure output
    print('-dpng',['LMAX',titlestr,'_',num2str(j),'.png']);
end
\end{verbatim}
\normalsize
\vskip\medskipamount 
\leaders\vrule width \textwidth\vskip0.4pt 
\vskip\medskipamount 
\nointerlineskip

\section{\label{sec:MATLABFunctions}MATLAB functions}

NB. The statistics toolbox is required only for the evaluation suite in \S \ref{sec:tdepdfsuite}.

\subsection{\label{sec:cv1d}cv1d.m}

\footnotesize
\begin{verbatim}
function cv = cv1d(X,kernelflag)

% Cross-validation bandwidth selector for 1D sample data X. Intended for
% comparison with tde1d. If sign(kernelflag) = 1, a Gaussian kernel is
% used; if sign(kernelflag) = -1, an Epanechnikov kernel is used
% 
% Output fields:
%     h,      bandwidth
%     x,      x-values of CV
%     y,      y-values of CV
%     a,      signficance levels
%     l,      lower bounds
%     u,      upper bounds

%% Preliminaries
X = X(:);
n = numel(X);
DX = max(X)-min(X);
if DX == 0
    warning('Dx = 0');
    cv.h = 0;
    cv.x = mean(X);
    cv.y = 1;
    cv.a = 1;
    cv.l = cv.y;
    cv.u = cv.y;
    return;
end

%% Select bandwidths and minimize risk
nh = min(n,100);
cv.h = NaN;
if kernelflag
    %% Select bandwidths
    h = DX./(1:nh);
    L = linspace(min(X),max(X),nh);
    %% Minimize risk
    risk0 = Inf;
    for i = 1:numel(h)  
        risk = cvrisk(X,h(i),kernelflag);
        if risk < risk0
            risk0 = risk;
            cv.h = h(i);
        end
    end
else
    error('kernelflag = 0: use lowriskhist.m instead');
end

%% Output
if kernelflag > 0	% Gaussian
    cv.x = L;   
    cv.y = zeros(1,numel(L));
    for k = 1:n
        cv.y = cv.y+exp(-.5*((L-X(k))/cv.h).^2)/(cv.h*sqrt(2*pi));
    end
    cv.y = cv.y/n;	% normalize
elseif kernelflag < 0	% Epanechnikov
    cv.x = L;   
    cv.y = zeros(1,numel(L));
    for k = 1:n
        cv.y = cv.y+(.75/cv.h)*max(0,1-((L-X(k))/cv.h).^2);
    end
    cv.y = cv.y/n;    % normalize
end

%% Compute (approximate) confidence bands 
% This would benefit from a running variance computation a la Knuth vol. 2,
% p. 232 or en.wikipedia.org/wiki/Algorithms_for_calculating_variance#Online_algorithm
d = ceil(log10(n));
cv.a = logspace(-1,-d,d);    % 1-(confidence levels)
cv.l = cell(1,d);  % lower bounds of approximate confidence bands
cv.u = cell(1,d);  % upper bounds of approximate confidence bands
if kernelflag 
    Y = zeros(n,numel(cv.x));
    % See section 20.3 of Wasserman's book All of Statistics 
    if kernelflag > 0	% Gaussian
        omega = 3*cv.h;
        for i = 1:n
            Y(i,:) = exp(-.5*((cv.x-X(i))/cv.h).^2)/(cv.h*sqrt(2*pi));
        end
    elseif kernelflag < 0	% Epanechnikov
        omega = 2*cv.h;
        for i = 1:n
            Y(i,:) = (.75/cv.h)*max(0,1-((cv.x-X(i))/cv.h).^2);
        end
    end
    m = DX/omega;
    q = erfinv((1-cv.a).^(1/m))/sqrt(2);
    s2 = var(Y,0,1);
    se = sqrt(s2/n);
    for j = 1:d
        cv.l{j} = cv.y-q(j)*se;
        cv.u{j} = cv.y+q(j)*se;
    end
else
    % See Theorem 20.10 of AoS and/or Theorem 6.20 of AoNS
    for j = 1:d
        z = erfinv(1-cv.a(j)/cv.h)/sqrt(2);
        c = sqrt(cv.h/n)*z/2;
        cv.l{j} = max(sqrt(cv.y)-c,0).^2;
        cv.u{j} = (sqrt(cv.y)+c).^2;
    end
end
\end{verbatim}
\normalsize
\vskip\medskipamount 
\leaders\vrule width \textwidth\vskip0.4pt 
\vskip\medskipamount 
\nointerlineskip

\subsection{\label{sec:cvrisk}cvrisk.m}

\footnotesize
\begin{verbatim}
function J_hat = cvrisk(X,h,kernelflag)

% Cross-validation estimate of risk for 1D data X and kernel bandwidth h.
% If sign(kernelflag) = 1, a Gaussian kernel is assumed; if
% sign(kernelflag) = -1, an Epanechnikov kernel is assumed; if kernelflag =
% 0, an error suggesting the use of an alternative function is returned.
%
% NB. The exact estimated risk is returned rather than a common (but quite
% accurate) approximation. The runtime penalty for this is negligible.
%
% NB. There are faster ways to compute the risk, at least for Gaussian
% kernels, using gridding/FFT a la Silverman or the fast Gauss transform.
% However, both of these have runtimes that depend on the precision desired
% and introduce a great deal of complexity into the process.

n = numel(X);
if kernelflag ~=0   % kernel
    X = X(:);
    XXh = bsxfun(@minus,X,X')/h;    % XXh(j,k) = (X(j)-X(k))/h
    % K is the kernel; K0 is its value at 0; K2 is the convolution of K
    % with itself
    if sign(kernelflag) == 1    % Gaussian
        XXh2 = XXh.^2;
        K0 = 1/sqrt(2*pi);
        K = exp(-(XXh2)/2)*K0;
        K2 = exp(-(XXh2)/4)*1/sqrt(4*pi);
    else    % Epanechnikov
        aXXh = abs(XXh);
        K0 = 3/4;
        K = (1-aXXh.^2).*(aXXh<=1)*K0;
        K2 = ((2-aXXh).^3).*(aXXh.^2+6*aXXh+4).*(aXXh<=2)*3/160;
    end
    Kn = K2-(2/(n-1))*(n*K-K0);
    J_hat = sum(Kn(:))/(h*n^2);
else    
    error('kernelflag = 0: use lowriskhist.m instead');
end
\end{verbatim}
\normalsize
\vskip\medskipamount 
\leaders\vrule width \textwidth\vskip0.4pt 
\vskip\medskipamount 
\nointerlineskip

\subsection{\label{sec:ISE}ISE.m}

\footnotesize
\begin{verbatim}
function ISE = ise(x,p,X,h,flag)

% For argument x of a PDF p and samples X drawn from p, this computes the
% integrated square error (ISE) of a kernel density estimate with bandwidth
% h. If flag > 0, the kernel is Gaussian; otherwise, the kernel is
% Epanechnikov.

X = X(:);
n = numel(X);

%% Form KDE with bandwidth h
if flag > 0	% Gaussian KDE
    p_hat = zeros(1,numel(x));
    for k = 1:n
        p_hat = p_hat+(1/n)*exp(-.5*((x-X(k))/h).^2)/(h*sqrt(2*pi));
    end
else	% Epanechnikov KDE
    p_hat = zeros(1,numel(x));
    for k = 1:n
        p_hat = p_hat+(1/n)*(.75/h)*max(0,1-((x-X(k))/h).^2);
    end        
end
%% Output
ISE = sum(((p(2:end)-p_hat(2:end)).^2).*diff(x));
\end{verbatim}
\normalsize
\vskip\medskipamount 
\leaders\vrule width \textwidth\vskip0.4pt 
\vskip\medskipamount 
\nointerlineskip

\subsection{\label{sec:layertrans2}layertrans2.m}

\footnotesize
\begin{verbatim}
function h = layertrans2(x,y,A,B)

% Transparency plot of two matrices. This particular code avoids annoyances
% that arise when trying to use surf with transparency options.

%% Basic assertions on input sizes
assert(all(size(A)==size(B)),'A and B incompatible');
assert(size(A,2)==numel(x)-1,'x mismatch');
assert(size(A,1)==numel(y)-1,'y mismatch');

%% Affine transformations to unit interval
A = affineunit(A);
B = affineunit(B);

%% Plot
hold on;
for j = 1:size(A,1)
    for k = 1:size(A,2)
        pA = patch([x(k),x(k+1),x(k+1),x(k)]+.5*(x(k+1)-x(k)),...
            [y(j),y(j),y(j+1),y(j+1)]+.5*(y(j+1)-y(j)),'b');
        set(pA,'FaceAlpha',A(j,k),'EdgeAlpha',0);
        pB = patch([x(k),x(k+1),x(k+1),x(k)]+.5*(x(k+1)-x(k)),...
            [y(j),y(j),y(j+1),y(j+1)]+.5*(y(j+1)-y(j)),'r');
        set(pB,'FaceAlpha',B(j,k),'EdgeAlpha',0);
    end
end
xlim(mean(x(1:2))+[min(x),max(x)]);
ylim(mean(y(1:2))+[min(y),max(y)]);

end

%% LOCAL FUNCTION

function A = affineunit(A)

minA = min(A(:));
maxA = max(A(:));
constA = (minA==maxA);  % 1 iff A is constant
A = (A-minA)/(maxA-minA+constA);

end
\end{verbatim}
\normalsize
\vskip\medskipamount 
\leaders\vrule width \textwidth\vskip0.4pt 
\vskip\medskipamount 
\nointerlineskip

\subsection{\label{sec:optimalkernelbandwidth}optimalkernelbandwidth.m}

\footnotesize
\begin{verbatim}
function y = optimalkernelbandwidth(x,f,n,kernelflag)

% For a PDF f(x), this gives the optimal bandwidth for a kernel density
% estimate and the approximate corresponding risk. See Wasserman's All of
% Statistics, Theorem 20.14. If kernelflag > 0, a Gaussian kernel is
% assumed; if kernelflag < 0, an Epanechnikov kernel is assumed; otherwise,
% an error is returned.
%
% WARNING: It is ASSUMED that x is regularly spaced and f is suitably nice.

%% Preliminaries
if numel(x) ~= numel(f)
    error('x and f are incompatibly sized');
else
    f = reshape(f,size(x));
end

%% c1, c2
if kernelflag > 0   % Gaussian
    c1 = 1;
    c2 = .5*sqrt(pi);
elseif kernelflag < 0   % Epanechnikov
    c1 = 1/5;
    c2 = 3/5;
else
    error('bad kernel flag');
end

%% c3
dx = mean(diff(x));
d2fdx2 = conv(f,[1,-2,1],'valid')/dx^2;
c3 = sum((d2fdx2.^2))*dx;

%% Output
y.h = (c1^-.4)*(c2^.2)*(c3^-.2)*(n^-.2);
y.R = .25*(1^4)*(y.h^4)*c3+c2*(y.h^-1)*(n^-1);
\end{verbatim}
\normalsize
\vskip\medskipamount 
\leaders\vrule width \textwidth\vskip0.4pt 
\vskip\medskipamount 
\nointerlineskip

\subsection{\label{sec:tde1d}tde1d.m}

\footnotesize
\begin{verbatim}
function tde = tde1d(X,kernelflag)

% One-dimensional topological density estimate of sample data X. If flag =
% 0, a histogram is returned; if sign(kernelflag) = 1, a Gaussian kernel
% density estimate is returned; if sign(kernelflag) = -1, an Epanechnikov
% kernel density estimate is returned.
% 
% Output fields:
%     h,      bandwidth
%     x,      x-values of TDE
%     y,      y-values of TDE
%     uc,     unimodal category
%     mfuc,   most frequent unimodal category
%     a,      signficance levels
%     l,      lower bounds
%     u,      upper bounds

% NB. The use of histc vs. histcounts is deliberate to accomodate legacy
% installations of MATLAB.

%% Preliminaries
X = X(:);
n = numel(X);
DX = max(X)-min(X);
if DX == 0
    warning('Dx = 0');
    tde.h = 0;
    tde.x = mean(X);
    tde.y = 1;
    tde.uc = 1;
    tde.mfuc = 1;
    tde.a = 1;
    tde.l = tde.y;
    tde.u = tde.y;
    return;
end

%% Select bandwidths/bin numbers
nh = min(n,100);
if kernelflag
    %% Select bandwidths
    h = DX./(1:nh);
else
    %% Select bin numbers
    h = 1:nh;
end
L = linspace(min(X),max(X),nh);

%% Compute unimodal category as a function of h
uc = zeros(1,nh);
for j = 1:nh
    %% Form density estimate corresponding to h(j)
    if kernelflag > 0	% Gaussian KDE
        p_hat = zeros(1,numel(L));
        for k = 1:n
            p_hat = p_hat+exp(-.5*((L-X(k))/h(j)).^2)/(h(j)*sqrt(2*pi));
        end
        p_hat = p_hat/n;	% normalize
    elseif kernelflag < 0 % Epanechnikov KDE
        p_hat = zeros(1,numel(L));
        for k = 1:n
            p_hat = p_hat+(.75/h(j))*max(0,1-((L-X(k))/h(j)).^2);
        end   
        p_hat = p_hat/n;	% normalize
    else	% histogram
        p_hat = histc(X,linspace(min(L),max(L),h(j)))/n;
    end        
    %% Compute unimodal category
    u = unidec(p_hat,0);
    uc(j) = numel(sum(u,2)>0);
end

%% Find most frequent unimodal category
uuc = unique(uc);
fuc = zeros(1,numel(uuc));
for j = 1:numel(uuc)
    fuc(j) = nnz(uc==uuc(j));
end
mfuc = uuc(find(fuc==max(fuc),1,'first'));

%% Find central number of bins for the most frequent unimodal category
temp = find(uc==mfuc);
h_opt = h(temp(ceil(numel(temp)/2)));
tde.h = h_opt;

%% Output
if kernelflag > 0	% Gaussian
    tde.x = L;   
    tde.y = zeros(1,numel(L));
    for k = 1:n
        tde.y = tde.y+exp(-.5*((L-X(k))/h_opt).^2)/(h_opt*sqrt(2*pi));
    end
    tde.y = tde.y/n;	% normalize
elseif kernelflag < 0	% Epanechnikov
    tde.x = L;   
    tde.y = zeros(1,numel(L));
    for k = 1:n
        tde.y = tde.y+(.75/h_opt)*max(0,1-((L-X(k))/h_opt).^2);
    end
    tde.y = tde.y/n;    % normalize
else    % histogram
    tde.x = linspace(min(L),max(L),h_opt);	% optimal bin centers
    tde.y = histc(X,tde.x)/n;               % corresp. normalized counts
end
tde.uc = uc;
tde.mfuc = mfuc;

%% Compute (approximate) confidence bands 
% This would benefit from a running variance computation a la Knuth vol. 2,
% p. 232 or en.wikipedia.org/wiki/Algorithms_for_calculating_variance#Online_algorithm
d = ceil(log10(n));
tde.a = logspace(-1,-d,d);    % 1-(confidence levels)
tde.l = cell(1,d);  % lower bounds of approximate confidence bands
tde.u = cell(1,d);  % upper bounds of approximate confidence bands
if kernelflag 
    Y = zeros(n,numel(tde.x));
    % See section 20.3 of Wasserman's book All of Statistics 
    if kernelflag > 0	% Gaussian
        omega = 3*tde.h;
        for i = 1:n
            Y(i,:) = exp(-.5*((tde.x-X(i))/tde.h).^2)/(tde.h*sqrt(2*pi));
        end
    elseif kernelflag < 0	% Epanechnikov
        omega = 2*tde.h;
        for i = 1:n
            Y(i,:) = (.75/tde.h)*max(0,1-((tde.x-X(i))/tde.h).^2);
        end
    end
    m = DX/omega;
    q = erfinv((1-tde.a).^(1/m))/sqrt(2);
    s2 = var(Y,0,1);
    se = sqrt(s2/n);
    for j = 1:d
        tde.l{j} = tde.y-q(j)*se;
        tde.u{j} = tde.y+q(j)*se;
    end
else
    % See Theorem 20.10 of AoS and/or Theorem 6.20 of AoNS
    for j = 1:d
        z = erfinv(1-tde.a(j)/h_opt)/sqrt(2);
        c = sqrt(h_opt/n)*z/2;
        tde.l{j} = max(sqrt(tde.y)-c,0).^2;
        tde.u{j} = (sqrt(tde.y)+c).^2;
    end
end
\end{verbatim}
\normalsize
\vskip\medskipamount 
\leaders\vrule width \textwidth\vskip0.4pt 
\vskip\medskipamount 
\nointerlineskip

\subsection{\label{sec:tdepdfsuite}tdepdfsuite.m}

\footnotesize
\begin{verbatim}
function [X,f] = tdepdfsuite(n_x,n)

% Generate samples X and PDFs f for evaluation of topological density
% estimation. n_x is the number of points to evaluate PDFs at; n is the
% number of samples for each PDF.
%
% NB. The statistics toolbox is required for this function.
%
% Example:
%   [X,f] = tdepdfsuite(500,200);
 
%% Preliminaries
% Linear space
x_0 = -1;
x_1 = 2;
x = linspace(x_0,x_1,n_x);
% Samples
X = [];

%% Laplace distribution $f_1$
j = size(X,1)+1;
% Parameters
mu = .5;
b = .125;
theta{j} = [mu,b];
% Sample data
obj = makedist('Uniform');
U = random(obj,1,n)-.5;
X(j,:) = theta{j}(1)-theta{j}(2)*sign(U).*log(1-2*abs(U));
% Explicit PDF
f(j,:) = .5*exp(-abs(x-theta{j}(1))/theta{j}(2))/theta{j}(2);

%% Simple gamma distribution $f_2$
j = size(X,1)+1;
% Parameters
b = 1.5;
a = b^2;    
ell = 5;
shape = a;
scale = 1/(ell*b);
theta{j} = [shape,scale];
% Sample data
obj = makedist('Gamma','a',shape,'b',scale);
X(j,:) = random(obj,1,n);
% Explicit PDF
f(j,:) = pdf(obj,x);

%% Mixture of three gamma distributions $f_3$
j = size(X,1)+1;
% Parameters
mix = ones(3,1)/3;
cmix = [0;cumsum(mix(1:end-1))];
b = [1.5;3;6];
a = b.^2;
ell = 8;
shape = a;
scale = 1./(ell*b);
theta{j} = [mix,shape,scale];
% Sample data
for i = 1:3
    obj(i) = makedist('Gamma','a',shape(i),'b',scale(i));
end
X(j,:) = zeros(1,n);
for k = 1:n
    i = find(cmix<rand,1,'last');
    X(j,k) = random(obj(i),1,1);
end 
% Explicit PDF
f(j,:) = zeros(1,n_x);
for i = 1:3
    f(j,:) = f(j,:)+mix(i)*pdf(obj(i),x);
end

%% Simple normal distribution $f_4$
j = size(X,1)+1;
% Parameters
mu = 0.5;
sigma = 0.2;
theta{j} = [mu,sigma];
% Sample data
obj = gmdistribution(mu,sigma'*sigma);
X(j,:) = random(obj,n)';
% Explicit PDF
f(j,:) = pdf(obj,x');

%% Mixture of two normal distributions $f_5$
j = size(X,1)+1;
% Parameters
mu = [0.35;0.65];
sigma = cat(3,0.1^2,0.1^2);
mix = ones(2,1)/2;
theta{j} = [mix,mu,[sigma(1);sigma(2)]];
% Sample data and explicit PDF
obj = gmdistribution(mu,sigma,mix);
X(j,:) = random(obj,n)';
% Explicit PDF
f(j,:) = pdf(obj,x');

%% Mixture of three normal distributions $f_6$
j = size(X,1)+1;
% Parameters
mu = [0.25;0.5;0.75];
sigma = cat(3,0.075^2,0.075^2,0.075^2);
mix = ones(3,1)/3;
theta{j} = [mix,mu,[sigma(1);sigma(2);sigma(3)]];
% Sample data and explicit PDF
obj = gmdistribution(mu,sigma,mix);
X(j,:) = random(obj,n)';
% Explicit PDF
f(j,:) = pdf(obj,x');

%% Mixtures of varying numbers of normal distributions $f_{km}$
for k = 1:3
    for m = 1:10
        j = size(X,1)+1;
        % Parameters
        mu = (1:m)'/(m+1);
        s2 = (2^-(k+2))*(m+1)^-2;
        sigma = reshape(s2*ones(1,m),[1,1,m]);
        mix = ones(m,1)/m;
        theta{j} = [mix,mu,reshape(sigma,[m,1])];
        % Sample data and explicit PDF
        obj = gmdistribution(mu,sigma,mix);
        X(j,:) = random(obj,n)';
        % Explicit PDF
        f(j,:) = pdf(obj,x');       
    end
end
\end{verbatim}
\normalsize
\vskip\medskipamount 
\leaders\vrule width \textwidth\vskip0.4pt 
\vskip\medskipamount 
\nointerlineskip

\subsection{\label{sec:unidec}unidec.m}

\footnotesize
\begin{verbatim}
function u = unidec(f,plotflag)

% Computes an explicit unimodal decomposition of a (possibly nonnormalized)
% PDF (i.e., a nonnegative function) f with bounded support. For details
% (and some otherwise cryptic notational choices), see Baryshnikov, Y. and
% Ghrist, R. "Unimodal category and topological statistics." NOLTA (2011).

%% Preliminaries
if any(f<0), error('f must be nonnegative'); end
f = reshape(f,[1,numel(f)]);
f = [0,f,0];
S = sum(f); 
f = f/S;    % Normalize to unit sum (reverted below)
N = numel(f);
% df = [0,diff(f)];

%% Sweep algorithm
% Step 1
alpha = 1;
u(alpha,:) = zeros(1,N);
g(alpha,:) = f;
% Steps 2/9
while any(g(alpha,:))
    % Step 3: y_alpha = first maximum of g(alpha,:) from left
    y_alpha = find(diff(g(alpha,:))<0,1,'first');
    L = 1:y_alpha;
    R = (y_alpha+1):N;
    % Step 4
    u(alpha+1,L) = g(alpha,L);
    % Step 5 (tweak here courtesy of Jeong-O Jeong)
    df = [0,diff(g(alpha,:))];
    du = min(df(R),0);
    u(alpha+1,R) = u(alpha+1,y_alpha)+cumsum(du);
    % Step 6
    u = max(u,0);
    % Step 7
    alpha = alpha+1;
    % Step 8: alternatively, g(alpha,:) = f-sum(u(1:alpha,:),1)
    g(alpha,:) = g(alpha-1,:)-u(alpha,:);
    g = max(g,0);   % for numerical stability
end

%% Eliminate spurious artifacts
% Revert normalization after elimination
u = S*u(sum(u,2)>eps,:);    
f = S*f;

%% Plot
if plotflag
    figure; 
    area(u')
    hold on
    plot(1:numel(f),f,'k')
    figure; 
    pcolorfull(log(u));
end
\end{verbatim}
\normalsize
\vskip\medskipamount 
\leaders\vrule width \textwidth\vskip0.4pt 
\vskip\medskipamount 
\nointerlineskip

\subsection{\label{sec:vbise}vbise.m}

\footnotesize
\begin{verbatim}
function y = vbise(x,p,X,flag)

% For argument x of a PDF p and samples X drawn from p, this computes the
% integrated square error (ISE) of kernel density estimates over a suitable
% set of bandwidths. If flag > 0, the kernel is Gaussian; otherwise, the
% kernel is Epanechnikov.

%% Select bandwidths
n = numel(X);
DX = max(X)-min(X);
if DX == 0, error('Dx = 0'); end
h = DX./(1:n);

%% Compute ISE as a function of h
ISE = zeros(1,n);
for j = 1:n
    ISE(j) = ise(x,p,X,h(j),flag);
end

%% Output
y.h = h;
y.ISE = ISE;
\end{verbatim}
\normalsize
\vskip\medskipamount 
\leaders\vrule width \textwidth\vskip0.4pt 
\vskip\medskipamount 
\nointerlineskip

\section{\label{sec:SmallSamples}Results for $n \in \{25,50,100\}$}

\begin{figure}[htbp]
\begin{tabular}{c c}
	Gaussian & Epanechnikov \\
	\includegraphics[width=90mm,keepaspectratio]{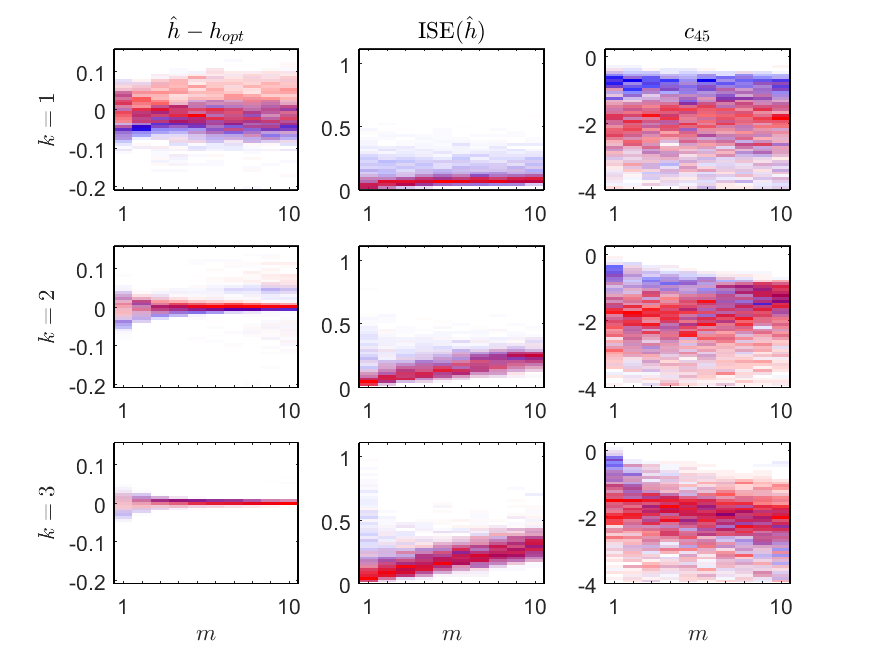}
	&
	\includegraphics[width=90mm,keepaspectratio]{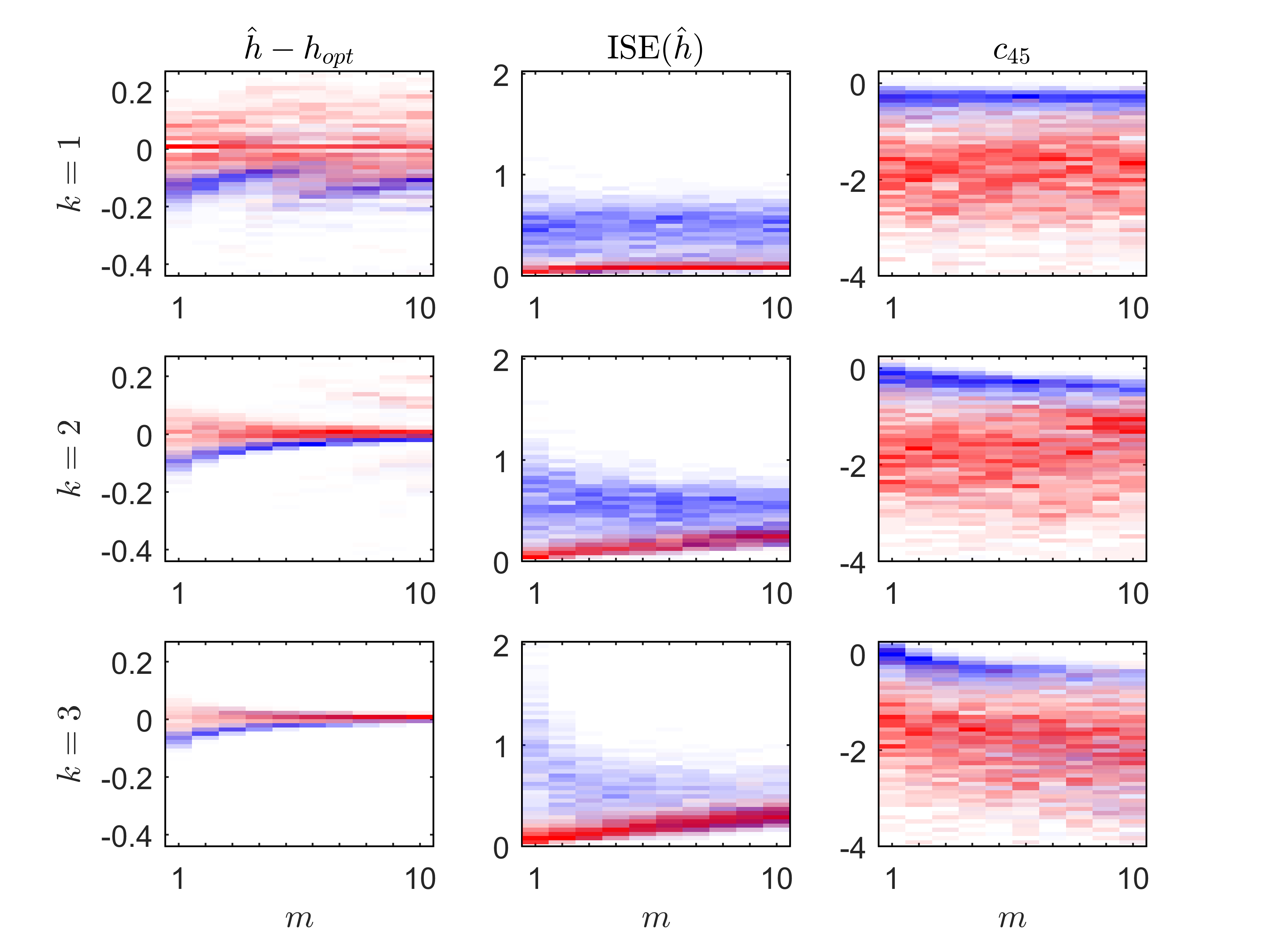}
\end{tabular}
\caption{ \label{fig:fkmEval100Matlab} As in figure \ref{fig:fkmEval500Matlab} for $n = 100$.}
\end{figure}

\begin{figure}[htbp]
\begin{tabular}{c c}
	Gaussian & Epanechnikov \\
	\includegraphics[width=90mm,keepaspectratio]{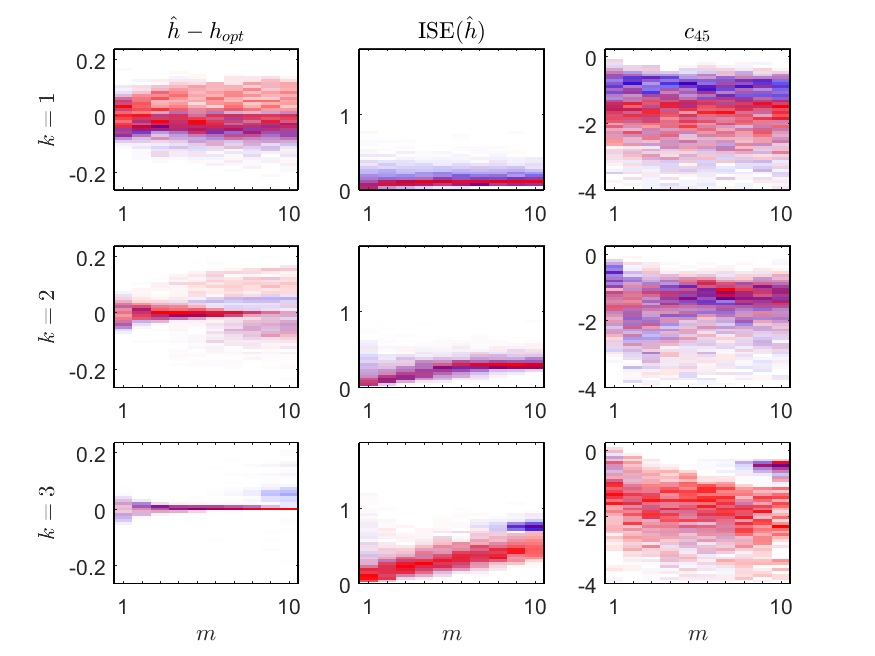}
	&
	\includegraphics[width=90mm,keepaspectratio]{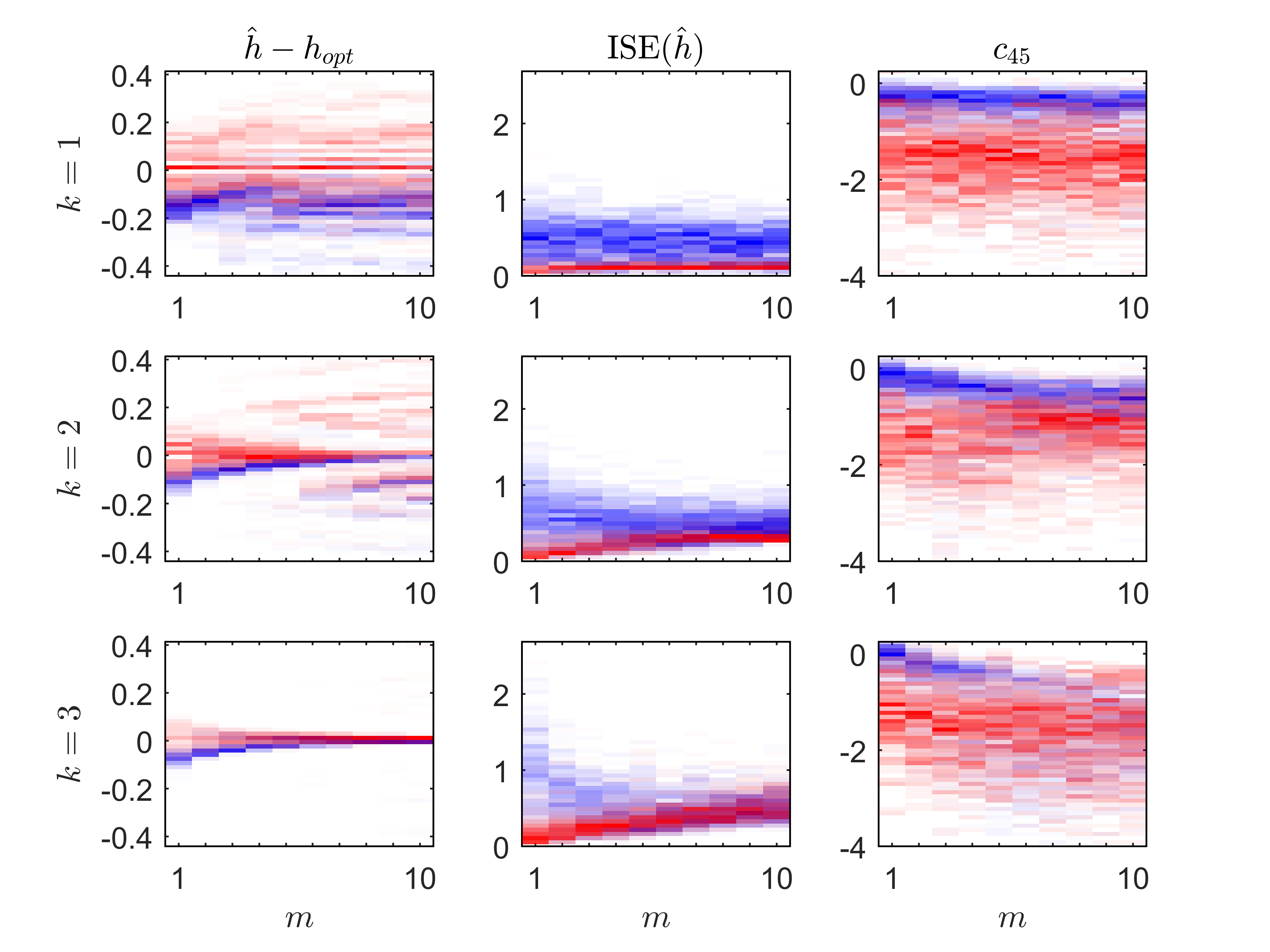}
\end{tabular}
\caption{ \label{fig:fkmEval50Matlab} As in figure \ref{fig:fkmEval500Matlab} for $n = 50$.}
\end{figure}

\begin{figure}[htbp]
\begin{tabular}{c c}
	Gaussian & Epanechnikov \\
	\includegraphics[width=90mm,keepaspectratio]{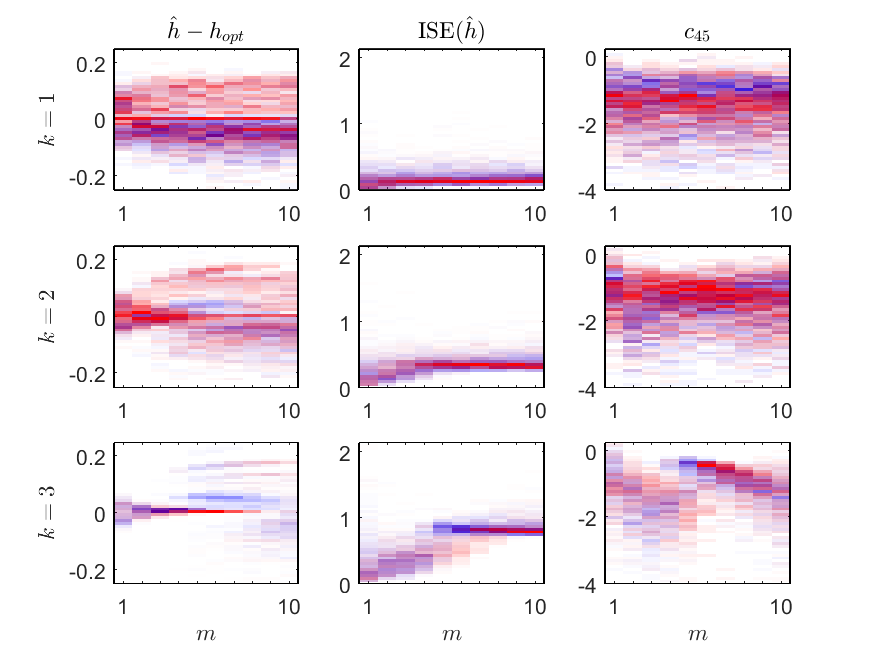}
	&
	\includegraphics[width=90mm,keepaspectratio]{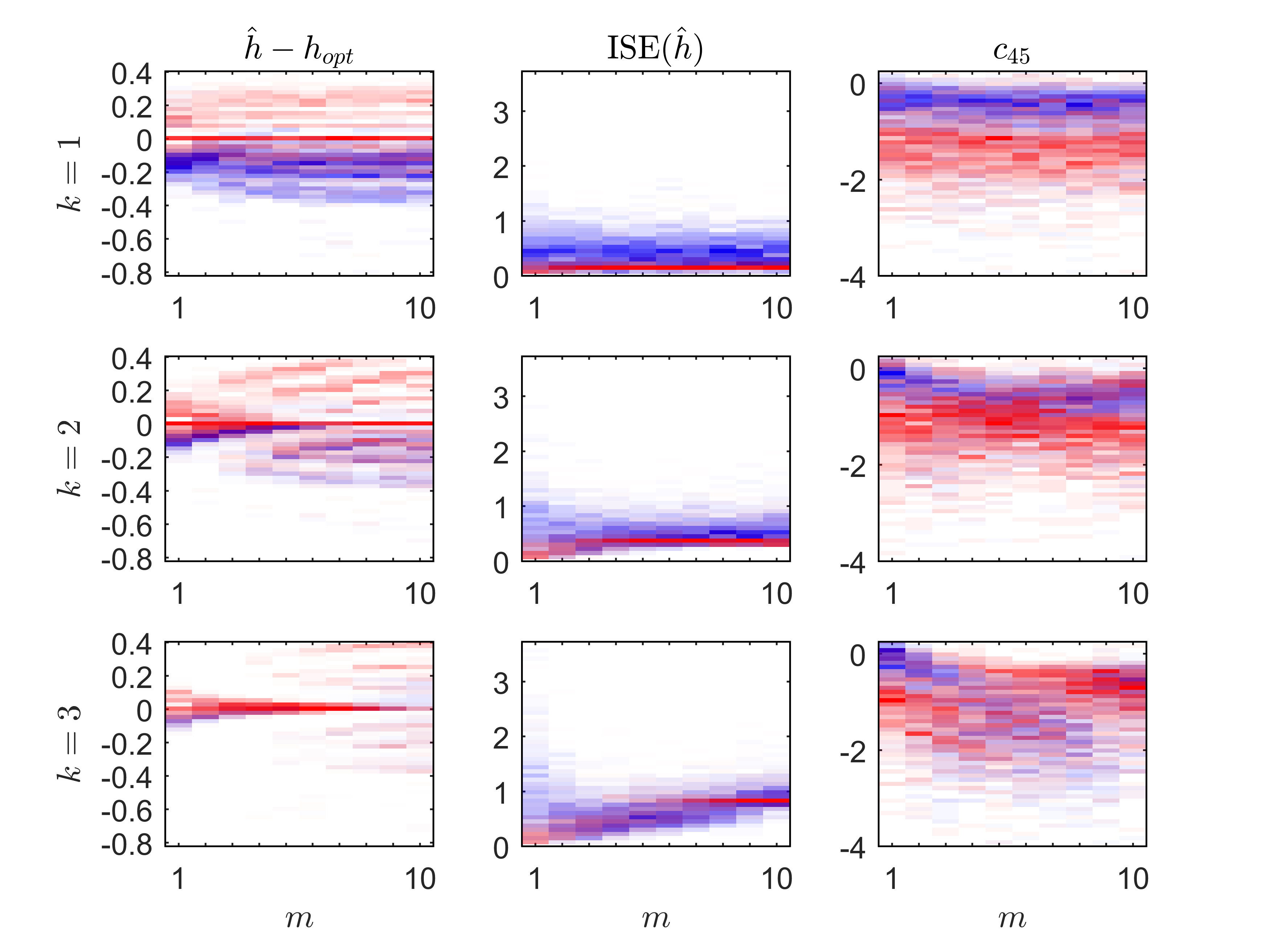}
\end{tabular}
\caption{ \label{fig:fkmEval25Matlab} As in figure \ref{fig:fkmEval500Matlab} for $n = 25$.}
\end{figure}

\begin{figure}[htbp]
\begin{tabular}{c c}
	Gaussian & Epanechnikov \\
	\includegraphics[width=90mm,keepaspectratio]{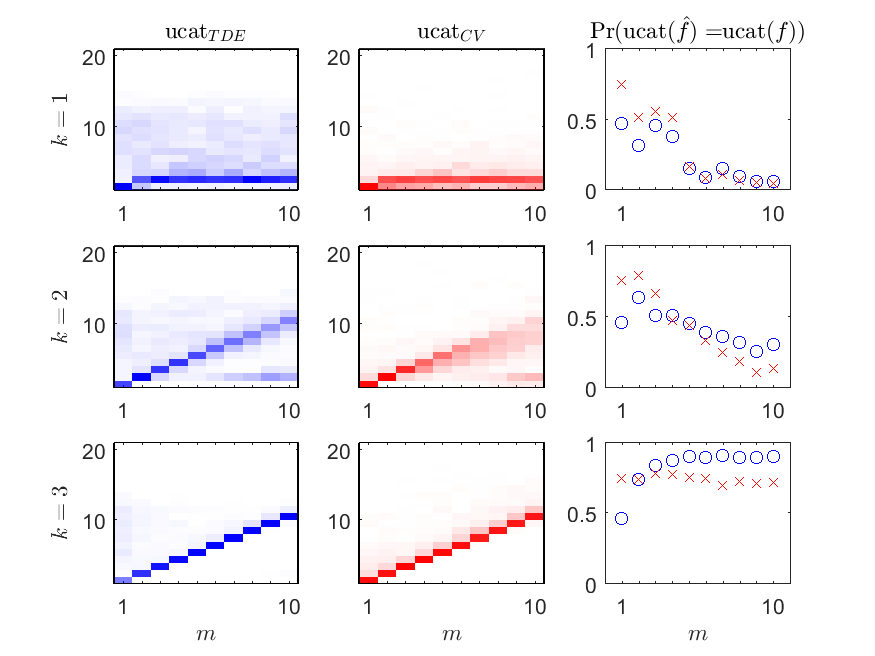}
	&
	\includegraphics[width=90mm,keepaspectratio]{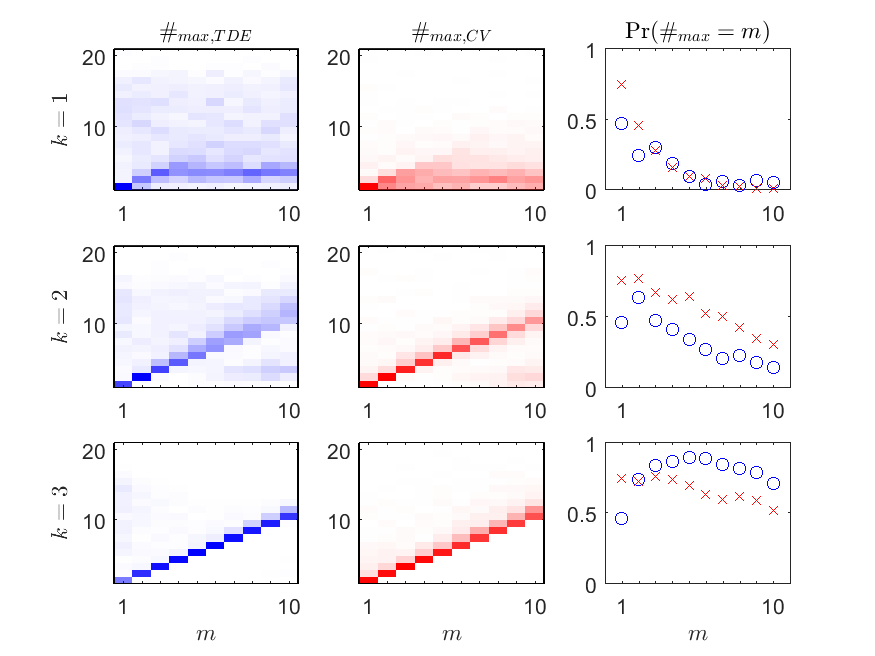}
\end{tabular}
\caption{ \label{fig:fkmEval100MatlabTop} As in figure \ref{fig:fkmEval500Matlab} for $n = 100$.}
\end{figure}

\begin{figure}[htbp]
\begin{tabular}{c c}
	Gaussian & Epanechnikov \\
	\includegraphics[width=90mm,keepaspectratio]{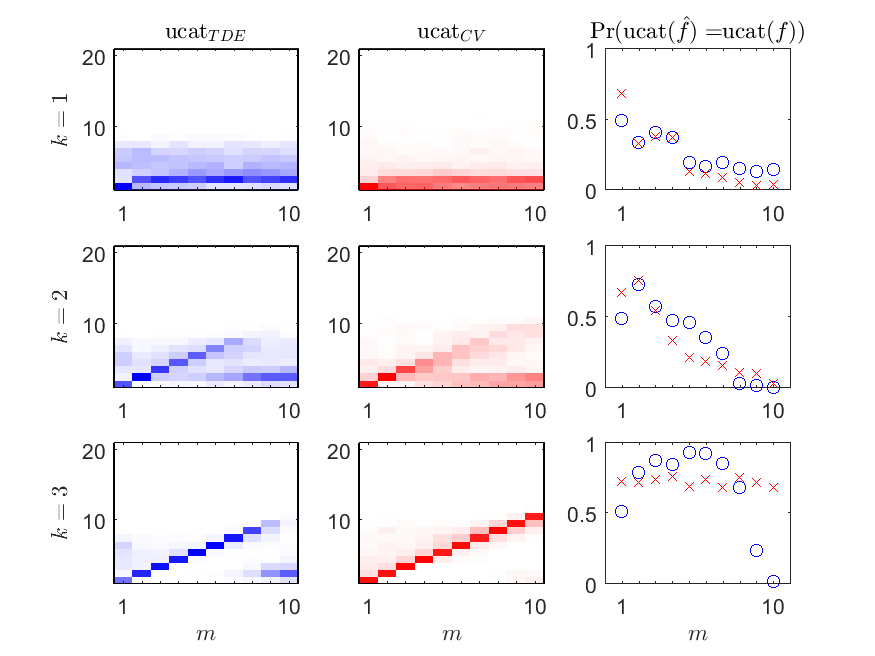}
	&
	\includegraphics[width=90mm,keepaspectratio]{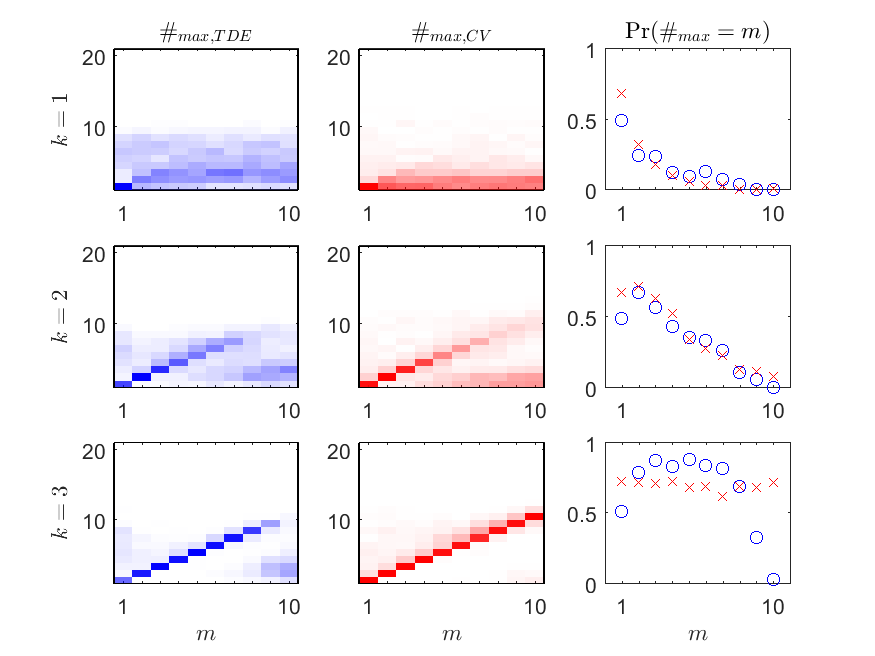}
\end{tabular}
\caption{ \label{fig:fkmEval50MatlabTop} As in figure \ref{fig:fkmEval500Matlab} for $n = 50$.}
\end{figure}

\begin{figure}[htbp]
\begin{tabular}{c c}
	Gaussian & Epanechnikov \\
	\includegraphics[width=90mm,keepaspectratio]{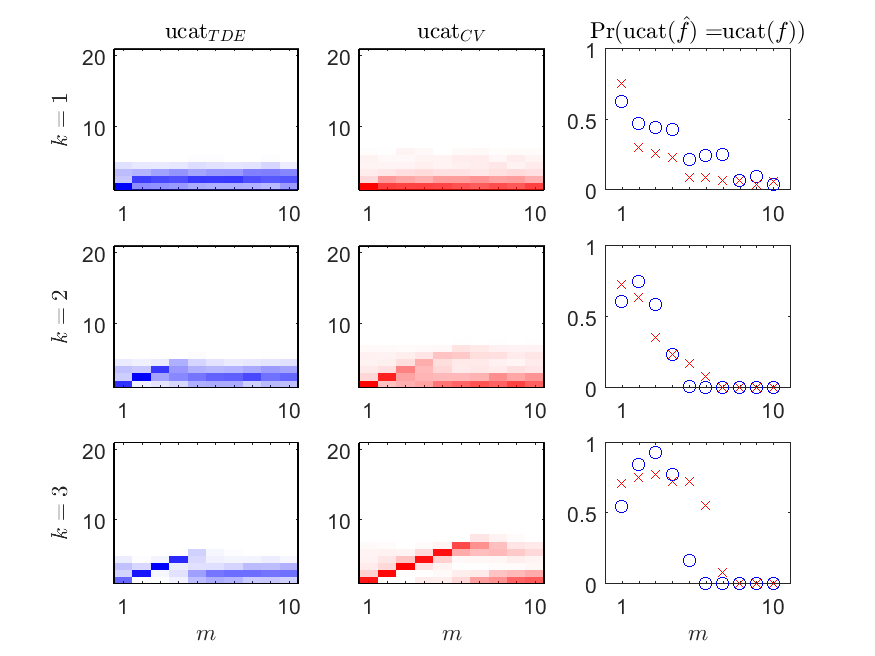}
	&
	\includegraphics[width=90mm,keepaspectratio]{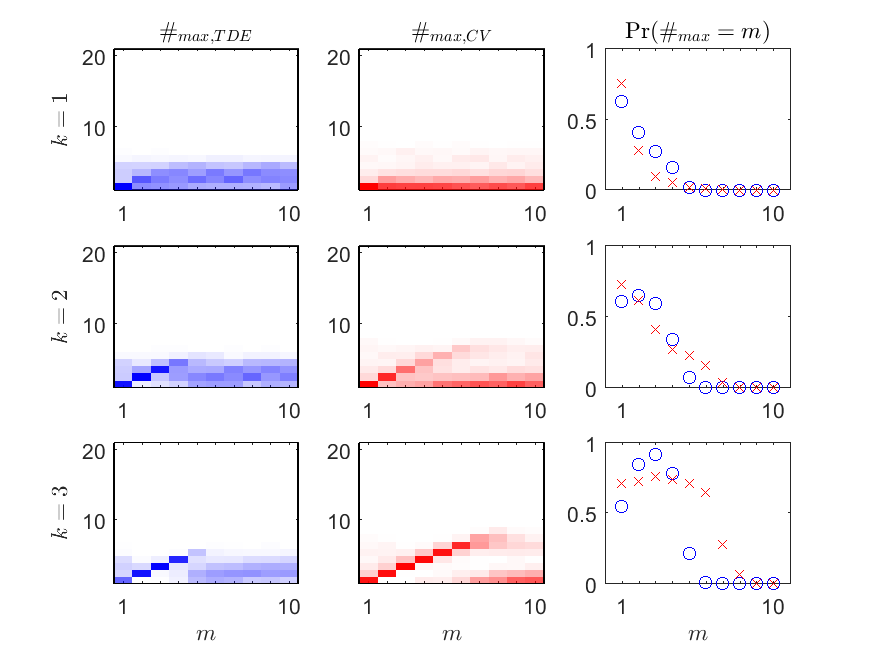}
\end{tabular}
\caption{ \label{fig:fkmEval25MatlabTop} As in figure \ref{fig:fkmEval500Matlab} for $n = 25$.}
\end{figure}

See figures \ref{fig:fkmEval100Matlab}-\ref{fig:fkmEval25MatlabTop}.

\section{\label{sec:fitdist} Comparison of TDE with MATLAB's \texttt{fitdist}}

\begin{figure}[htbp]
\begin{tabular}{c c}
	Gaussian & Epanechnikov \\
	\includegraphics[width=90mm,keepaspectratio]{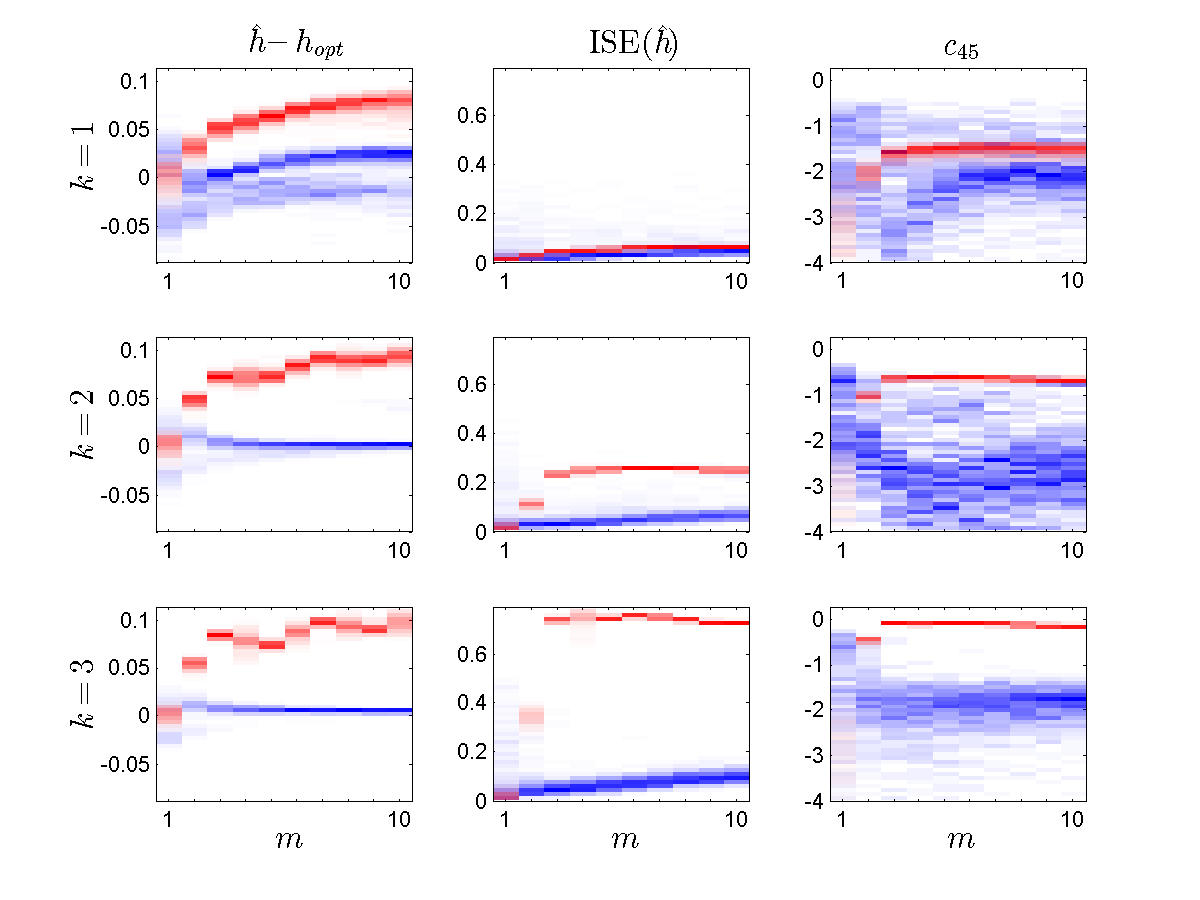}
	&
	\includegraphics[width=90mm,keepaspectratio]{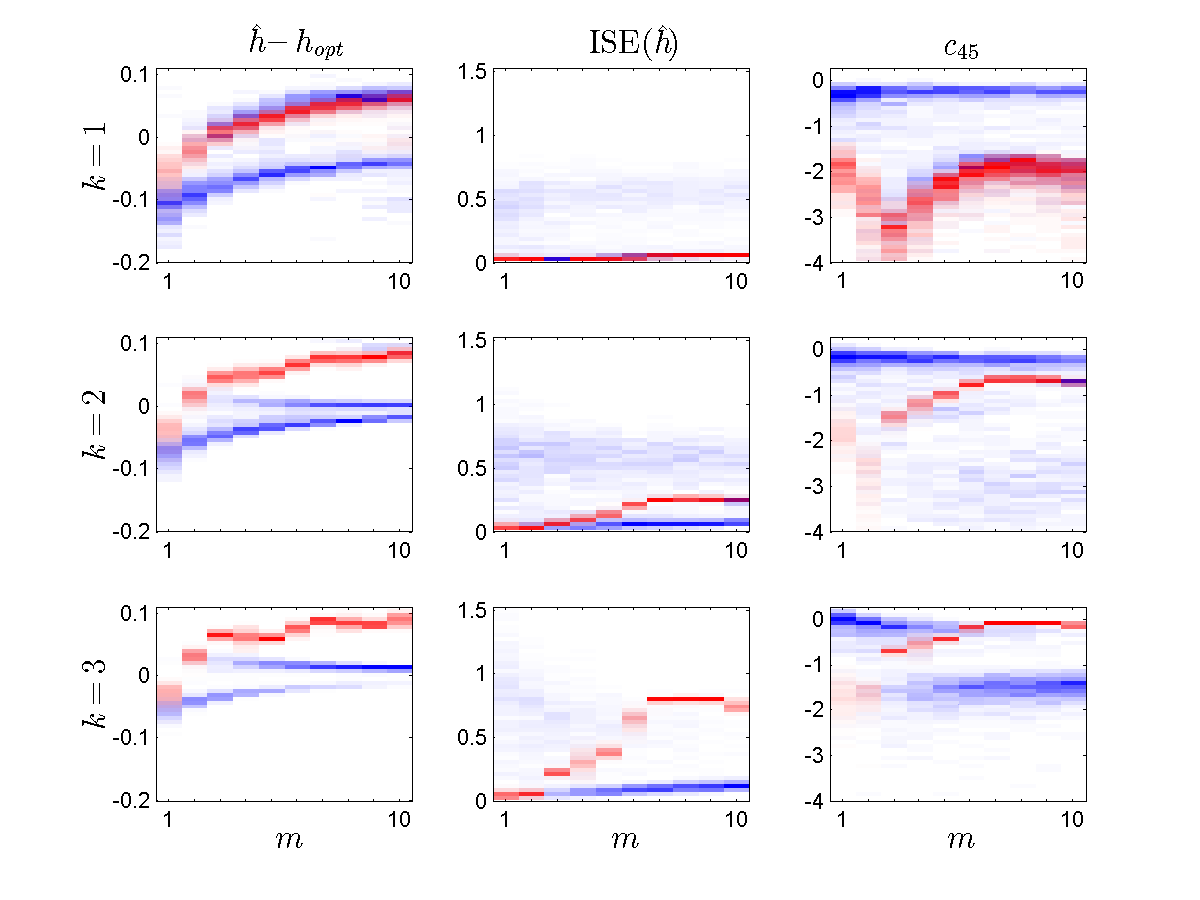}
\end{tabular}
\caption{ \label{fig:fkmEval500MatlabFitdist} As in figure \ref{fig:fkmEval500Matlab} for $n = 500$, but with {\color{red}red indicating data for MATLAB's \texttt{fitdist}} instead of CV.}
\end{figure}

\begin{figure}[htbp]
\begin{tabular}{c c}
	Gaussian & Epanechnikov \\
	\includegraphics[width=90mm,keepaspectratio]{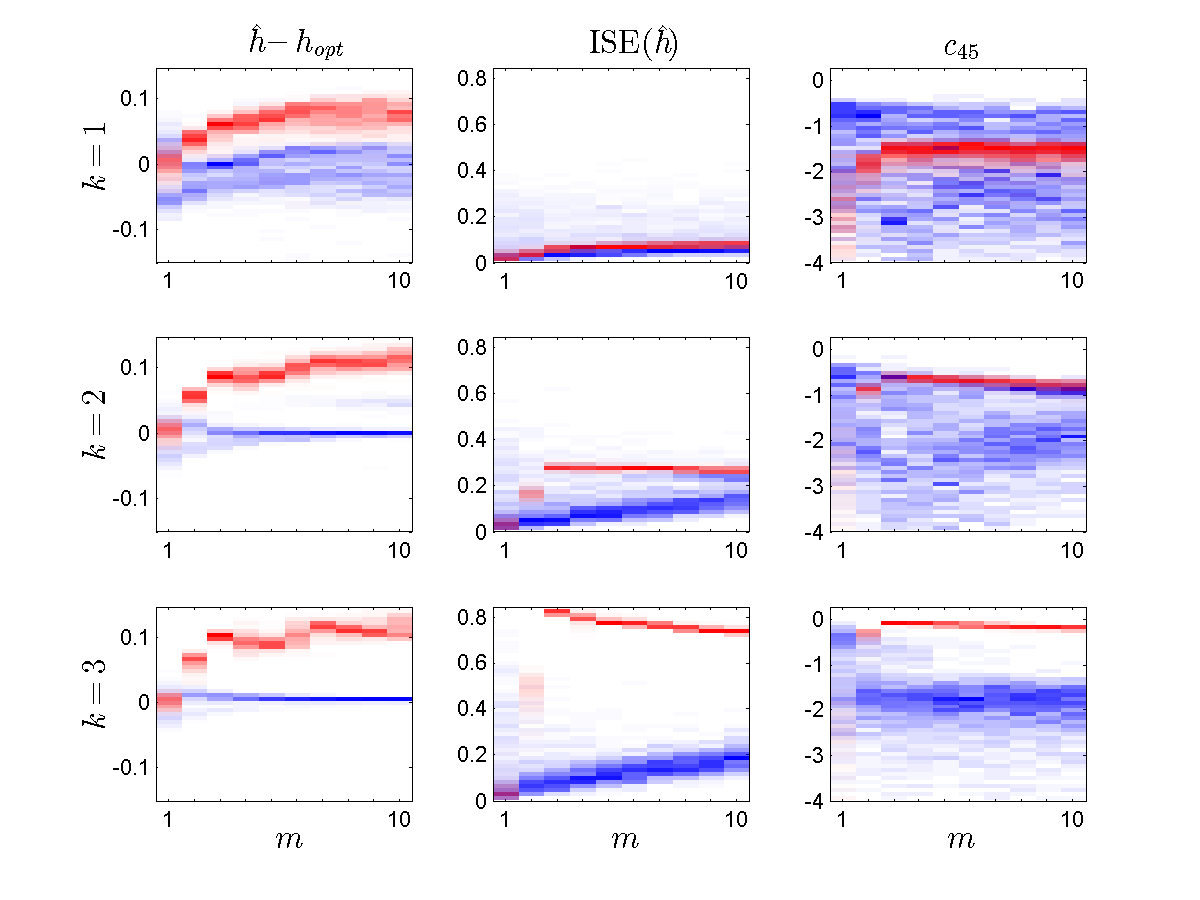}
	&
	\includegraphics[width=90mm,keepaspectratio]{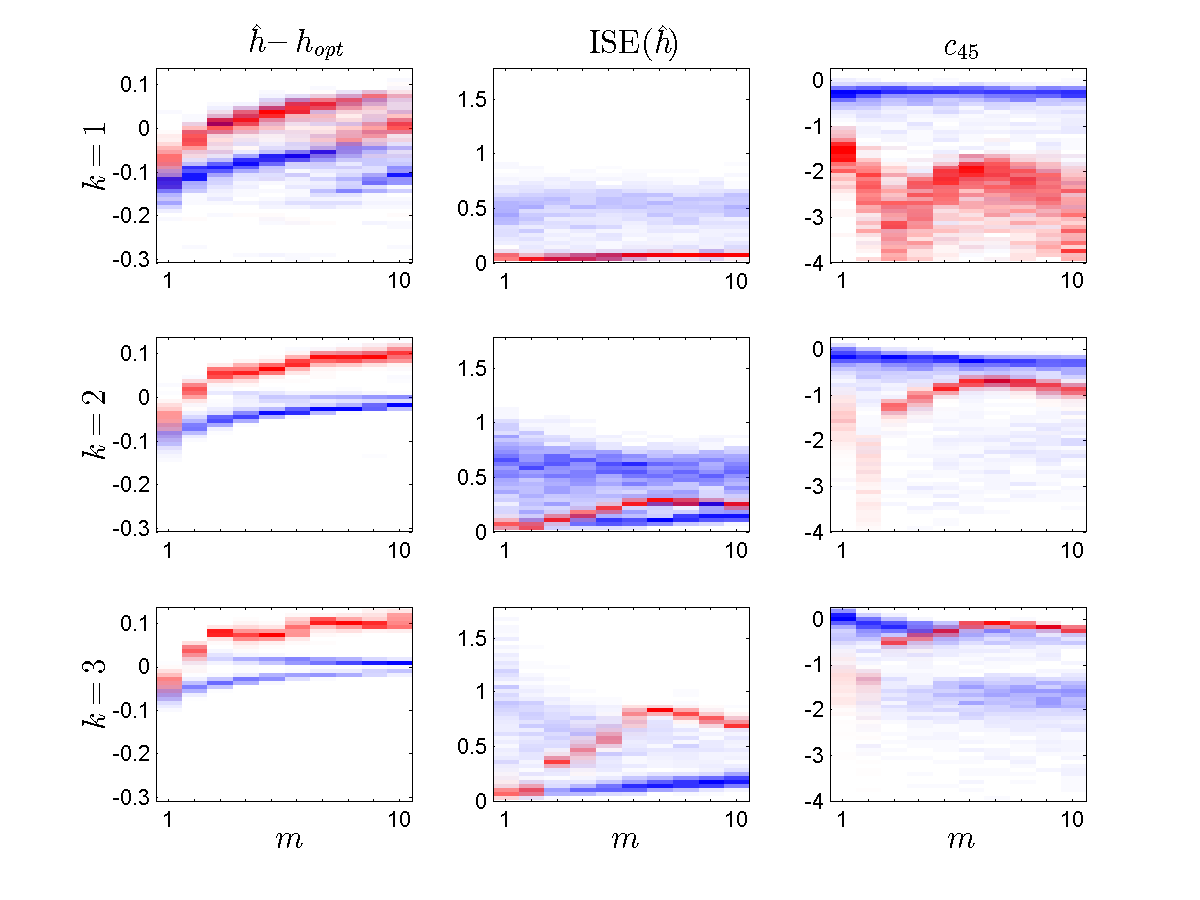}
\end{tabular}
\caption{ \label{fig:fkmEval200MatlabFitdist} As in figure \ref{fig:fkmEval500MatlabFitdist} for $n = 200$.}
\end{figure}

See figures \ref{fig:fkmEval500MatlabFitdist}-\ref{fig:fkmEval200MatlabFitdist}.

\begin{thebibliography}{16}

\bibitem{AndersonEtAl}Anderson, J. \emph{et al.} ``The more, the merrier: the blessing of dimensionality for learning large Gaussian mixtures.'' \emph{JMLR Workshop Conf. Proc.} {\bf 35}, 1 (2014).

\bibitem{BaryshnikovGhrist}Baryshnikov, Y. and Ghrist, R. ``Unimodal category and topological statistics.'' NOLTA (2011).

\bibitem{BauerEtAl}Bauer, U., \emph{et al.} ``Persistence barcodes versus Kolmogorov signatures: detecting modes of one-dimensional signals.'' \emph{Found. Comput. Math.} 1 (2015).

\bibitem{CarrSnoeyinkAxen}Carr, H., Snoeyink, J., and Axen, U. ``Computing contour trees in all dimensions.'' \emph{Comp. Geom.} {\bf 24}, 75 (2003).

\bibitem{ChaudhuriMarron}Chaudhuri, P. and Marron, J. S. ``Scale space view of curve estimation.'' \emph{Ann. Stat.} {\bf 28}, 402 (2000).

\bibitem{DevroyeLugosi}Devroye, L. and Lugosi, G. \emph{Combinatorial Methods in Density Estimation.} Springer (2001).

\bibitem{Ghrist}Ghrist, R. \emph{Elementary Applied Topology.} CreateSpace (2014). 

\bibitem{HallMinnotteZhang}Hall, P., Minotte, M. C., and Zhang, C. ``Bump hunting with non-Gaussian kernels.'' \emph{Ann. Stat.} {\bf 32}, 2124 (2004).

\bibitem{HeidenreichSchindlerSperlich}Heidenreich, N.-B., Schindler, A., and Sperlich, S. ``Bandwidth selection for kernel density estimation: a review of fully automatic selectors.'' \emph{Adv. Stat. Anal.} {\bf 97}, 403 (2013).

\bibitem{Minnotte}Minotte, M. C. ``Nonparametric testing of the existence of modes.'' \emph{Ann. Stat.} {\bf 25}, 1646 (1997).

\bibitem{Oudot}Oudot, S. Y. \emph{Persistence Theory: From Quiver Representations to Data Analysis.} AMS (2015).

\bibitem{PokornyEtAl1}Pokorny, F. T., \emph{et al.} ``Persistent homology for learning densities with bounded support.'' NIPS (2012).

\bibitem{PokornyEtAl2}Pokorny, F. T., \emph{et al.} ``Topological constraints and kernel-based density estimation.'' NIPS (2012).

\bibitem{Silverman}Silverman, B. W. ``Using kernel density estimation to investigate multimodality.'' \emph{J. Roy. Stat. Soc. B} {\bf 43}, 97 (1981).

\bibitem{Tsybakov}Tsybakov, A. B. \emph{Introduction to Nonparametric Estimation.} Springer (2009). 

\bibitem{Wasserman}Wasserman, L. \emph{All of Statistics: A Concise Course in Statistical Inference.} Springer (2004).

\bibitem{ZambomDias}Zambom, A. Z. and Dias, R. ``A review of kernel density estimation with applications to econometrics.'' \emph{Int. Econometric Rev.} {\bf 5}, 20 (2013).

\end{thebibliography}
\end{document}